\documentclass[aps,prl,twocolumn,secnumarabic,amsmath,amssymb,superscriptaddress]{revtex4-1}
\usepackage{amssymb,amsbsy,graphicx,subfigure,bm,multirow,epstopdf}
\usepackage{color}
\usepackage[colorlinks=true,
linkcolor=blue,
urlcolor=blue,
citecolor=blue]{hyperref}

\begin{document}

\title{Vison Crystals in an Extended Kitaev Model on the Honeycomb Lattice}

\author{Shang-Shun~Zhang}

\affiliation{Department of Physics and Astronomy, The University of
Tennessee, Knoxville, Tennessee 37996, USA}

\author{Zhentao~Wang}

\affiliation{Department of Physics and Astronomy, The University of
Tennessee, Knoxville, Tennessee 37996, USA}

\author{G{\'a}bor~B.~Hal{\'a}sz}

\affiliation{Materials Science and Technology Division, Oak Ridge
National Laboratory, Oak Ridge, Tennessee 37831, USA}

\author{Cristian~D.~Batista}

\affiliation{Department of Physics and Astronomy, The University of
Tennessee, Knoxville, Tennessee 37996, USA}

\affiliation{Neutron Scattering Division and Shull-Wollan Center,
Oak Ridge National Laboratory, Oak Ridge, Tennessee 37831, USA}

\begin{abstract}

We introduce an extension of the Kitaev honeycomb model by including
four-spin interactions that preserve the local gauge structure and
hence the integrability of the original model. The extended model
has a rich phase diagram containing five distinct vison crystals, as
well as a symmetric $\pi$-flux spin liquid with a Fermi surface of
Majorana fermions and a sequence of Lifshitz transitions. We discuss
possible experimental signatures and, in particular, present
finite-temperature Monte Carlo calculations of the specific heat and
the static vison structure factor. We argue that our extended model
emerges naturally from generic perturbations to the Kitaev honeycomb
model.

\end{abstract}

\pacs{~}

\maketitle

{\it Introduction.} The famous Kitaev model on the honeycomb lattice
\cite{kitaev2006anyons} is an exactly solvable yet experimentally
realistic model of a quantum spin liquid. In contrast to more
conventional magnetic phases, quantum spin liquids retain extensive
(quantum) fluctuations all the way down to zero temperature
\cite{balents2010spin}, where the spins appear to fractionalize into
deconfined ``spinon'' quasiparticles coupled to appropriate gauge
fields \cite{savary2016quantum}.

The Kitaev model is approximately realized in a family of strongly
spin-orbit-coupled honeycomb materials, where its anisotropic spin
interactions emerge between effective $J = 1/2$ angular momenta in
the $t_{2g}$ orbitals of $4d$ or $5d$ ions \cite{jackeli2009mott,
rau2016spin, trebst2017kitaev, hermanns2018physics}. To determine
the most accurate microscopic spin models for these Kitaev
materials, including (Na,Li)$_2$IrO$_3$
\cite{singh2010antiferromagnetic, liu2011longrange,
singh2012relevance, choi2012spin, ye2012direct, comin2012novel,
chun2015direct, williams2016incommensurate, kitagawa2018spin} and
$\alpha$-RuCl$_3$ \cite{plumb2014spin, sandilands2015scattering,
sears2015magnetic, majumder2015anisotropic, johnson2015monoclinic,
sandilands2016spin, banerjee2016proximate, sears2017phase,
banerjee2017neutron, baek2017evidence, do2017majorana,
banerjee2018excitations, hentrich2018unusual, kasahara2018majorana},
various extensions of the Kitaev model have been considered and
analyzed with a wide range of techniques \cite{chaloupka2010kitaev,
jiang2011possible, reuther2011finite, price2012critical,
rau2014generic, yamaji2014first, sizyuk2014importance,sela2014,
rousochatzakis2015phase, kim2015, yadav2016, kim2016,
winter2016challenges, kim2016quasimolecular, janssen2016honeycomb,
Hou2017, winter2017models, Samarakoon2017, Ran2017, Samarakoon2018,
gordon2019theory}. While these models are experimentally realistic
and have rich phase diagrams in the classical limit, it is
challenging to identify and characterize quantum phases in them. For
a start, the honeycomb lattice may harbor many different quantum
spin liquids \cite{lu2011z2, you2012doping}, and the Kitaev spin
liquid, captured by the Kitaev model, is only one among these many
candidates. In addition, a quantum spin liquid may also remain
``hidden'' by appearing on top of classical symmetry-breaking order
\cite{savary2012coulombic}.

From a more phenomenological point of view, the low-energy physics
of the Kitaev spin liquid is described by Majorana fermions
(spinons) with Dirac nodes, coupled to an emergent $\mathbb{Z}_2$
gauge field \cite{kitaev2006anyons}. At each plaquette of the
honeycomb lattice, the $\mathbb{Z}_2$ gauge field may form a $\pi$
flux, corresponding to a ``vison'' excitation. In turn, the presence
of such a vison affects the kinetic energy of the spinons via the
Berry phase $\pi$ picked up by each spinon moving around it. For the
pure Kitaev model, the spinons are governed by a nearest-neighbor
hopping problem (cf.~electrons in graphene) and, due to the lack of
frustration, the ground state has no visons at any plaquettes
\cite{kitaev2006anyons, Lieb94}. However, if the hopping problem is
frustrated by competing hopping amplitudes, the presence of a vison
may reduce the frustration and thus lower the kinetic energy of the
spinons. Such a frustration in the hopping amplitudes is known to
stabilize crystals of topological solitons, such as baby skyrmions
or merons, in itinerant magnets \cite{Ozawa16, Batista16, Ozawa17},
and one may thus expect it to stabilize analogous vison crystals in
the Kitaev spin liquid.

\begin{figure}[tbp]
\centering
\includegraphics[width=1.0\columnwidth]{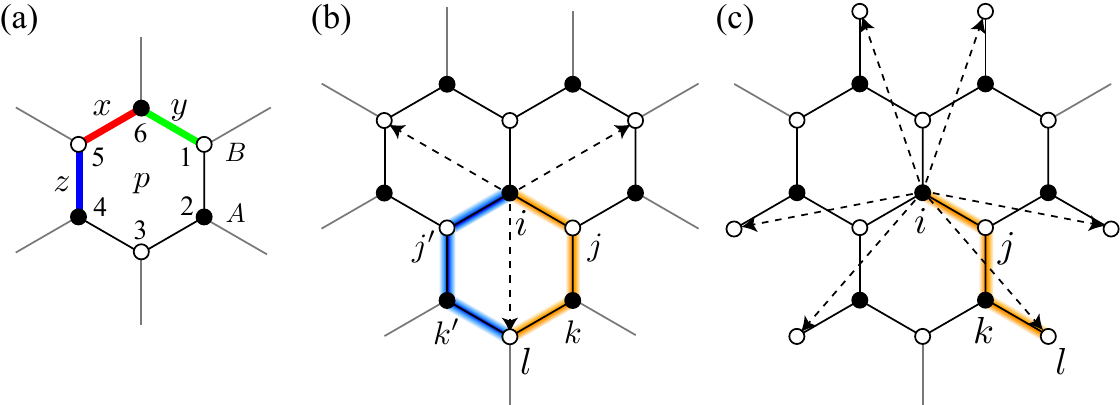}
\caption{Extended Kitaev model. (a) Honeycomb lattice with two
sublattices $A$ and $B$ (black and white dots), three bond types
$x$, $y$, and $z$ (red, green, and blue bonds), and the
site-labeling convention around a plaquette $p$. (b)-(c)
Representative (orange) paths $\langle ijkl \rangle_{yzx}$ (b) and
$\langle ijkl \rangle_{yzy}$ (c) associated with the $K_3$ and
$K_3'$ terms in Eq.~(\ref{eq:fourspin}), respectively; four-spin
interactions along such paths give rise to Majorana hopping from any
site $i$ to all its third neighbors \cite{note0}, as indicated by
the dashed arrows. For the path $\langle ijkl \rangle_{yzx}$ (b),
the symmetry-related path $\langle ij'k'l \rangle_{xzy}$ is marked
by blue.} \label{fig:model}
\end{figure}

In this Letter, we extend the Kitaev model by including four-spin
interactions that preserve the exact solution of the model and
emerge naturally from generic perturbations. By introducing
frustrated further-neighbor hopping for the Majorana fermions, these
additional interactions stabilize a rich variety of vison crystals,
as well as a symmetric $\pi$-flux spin liquid with a vison at every
plaquette. Interestingly, the $\pi$-flux spin liquid exhibits a
Fermi surface of Majorana fermions undergoing two subsequent
Lifshitz transitions. On a technical level, we first use a simple
variational treatment to compute the zero-temperature phase diagram
of our extended model. The validity of this approach is then
confirmed by unbiased Monte Carlo (MC) simulations that also reveal
the finite melting temperatures of the vison crystals.

{\it Model.} We consider a generalized Kitaev Hamiltonian on the honeycomb
lattice:
\begin{equation}
\label{eq:model} {\cal H} = {\cal H}_{K_1} + {\cal H}_{K_3},
\end{equation}
where ${\cal H}_{K_1} = - K_1 \sum_{\langle ij \rangle_{\alpha}}
\sigma^{\alpha}_i \sigma^{\alpha}_j$ is the usual
\cite{kitaev2006anyons} isotropic Kitaev Hamiltonian with
ferromagnetic ($K_1>0$) Ising interactions between the spin
components $\sigma^{\alpha}$ along each $\alpha = \{ x,y,z \}$ bond
$\langle ij \rangle_{\alpha}$ [see Fig.~\ref{fig:model}(a)], and
\begin{equation}
{\cal H}_{K_3} = K_3 \sum_{\langle
ijkl \rangle_{\alpha \beta \gamma}} \sigma_{i}^{\alpha}
\sigma_{j}^{\gamma} \sigma_{k}^{\alpha} \sigma_{l}^{\gamma}
- K_3^{\prime}  \sum_{\langle
ijkl \rangle_{\alpha \beta \alpha}} \sigma_{i}^{\alpha}
\sigma_{j}^{\gamma} \sigma_{k}^{\gamma} \sigma_{l}^{\alpha},
\label{eq:fourspin}
\end{equation}
where $(\alpha \beta \gamma)$ is a permutation of $(xyz)$ in each
term, and $\langle ijkl \rangle_{\alpha \beta \gamma}$ is a path of
length $3$ consisting of bonds $\langle ij \rangle_{\alpha}$,
$\langle jk \rangle_{\beta}$, and $\langle kl \rangle_{\gamma}$.
Each term in ${\cal H}_{K_3}$ is the product of the three terms in
${\cal H}_{K_1}$ that correspond to the three bonds along the
appropriate path. Different $K_3$ and $K_3^{\prime}$ terms are
related by space-group symmetries, simultaneously transforming the
lattice and the spins; particular examples of their respective
paths, with $(\alpha \beta \gamma) = (yzx)$, are depicted in
Figs.~\ref{fig:model}(b) and \ref{fig:model}(c). We remark that, for
each path $\langle ijkl \rangle_{\alpha \beta \gamma}$ going around
one ``half'' of a hexagon, connecting opposite vertices $i$ and $l$,
there is a symmetry-related path $\langle lk'j'i \rangle_{\alpha
\beta \gamma} = \langle ij'k'l \rangle_{\gamma \beta \alpha}$ going
around the other ``half'' of the hexagon [see
Fig.~\ref{fig:model}(b)].

Importantly, the exact solution of ${\cal H}_{K_1}$
\cite{kitaev2006anyons} is preserved by the additional terms in
Eq.~(\ref{eq:fourspin}). Indeed, since ${\cal H}$ commutes with the
flux operator $W_p = \sigma_1^x \sigma_2^y \sigma_3^z \sigma_4^x
\sigma_5^y \sigma_6^z$ at each plaquette $p$ [see
Fig.~\ref{fig:model}(a)], one can identify static $\mathbb{Z}_2$
flux or ``vison'' degrees of freedom at these plaquettes, each being
present (absent) if the corresponding $W_p$ takes eigenvalue $-1$
($+1$). Following the Majorana fermionization $\sigma_j^{\alpha} = i
b^{\alpha}_j c_j^{\phantom{\dag}}$, the Hamiltonian takes the form
\cite{supp}
\begin{align}
{\cal H} &= i K_1 \sum_{\langle ij \rangle_{\alpha}} u_{ij}^{\alpha}
\, c_i^{\phantom{\dag}} c_j^{\phantom{\dag}} + i K_{3} \sum_{\langle
ijkl \rangle_{\alpha \beta \gamma}} u_{ij}^{\alpha} u_{kj}^{\beta}
u_{kl}^{\gamma} \, c_i^{\phantom{\dag}}
c_l^{\phantom{\dag}} \nonumber \\
& \quad + i K_{3}^{\prime} \sum_{\langle ijkl \rangle_{\alpha \beta
\alpha}} u_{ij}^{\alpha} u_{kj}^{\beta} u_{kl}^{\alpha} \,
c_i^{\phantom{\dag}} c_l^{\phantom{\dag}}, \label{eq:Hextended}
\end{align}
where $u_{ij}^{\alpha} = -u_{ji}^{\alpha} \equiv i b_i^{\alpha}
b_j^{\alpha}$ is a $\mathbb{Z}_2$ gauge field along the $\alpha$
bond $\langle ij \rangle_{\alpha}$. Since these gauge fields are
conserved quantities, $u_{ij}^{\alpha} = \pm 1$, providing a
redundant description of the conserved gauge fluxes, $W_p = u_{12}^z
u_{32}^x u_{34}^y u_{54}^z u_{56}^x u_{16}^y = \pm 1$,
Eq.~(\ref{eq:Hextended}) is quadratic in the Majorana fermions
$c_i$, thus giving rise to free fermion (``spinon'') excitations
after a straightforward diagonalization~\cite{supp}. From the
perspective of the Majorana fermions, the $K_1$ terms describe
first-neighbor hopping, while the additional $K_3$ and $K_3'$ terms
describe third-neighbor hopping~\cite{note0}.

\begin{figure}[tbp]
\centering
\includegraphics[width=1.0\columnwidth]{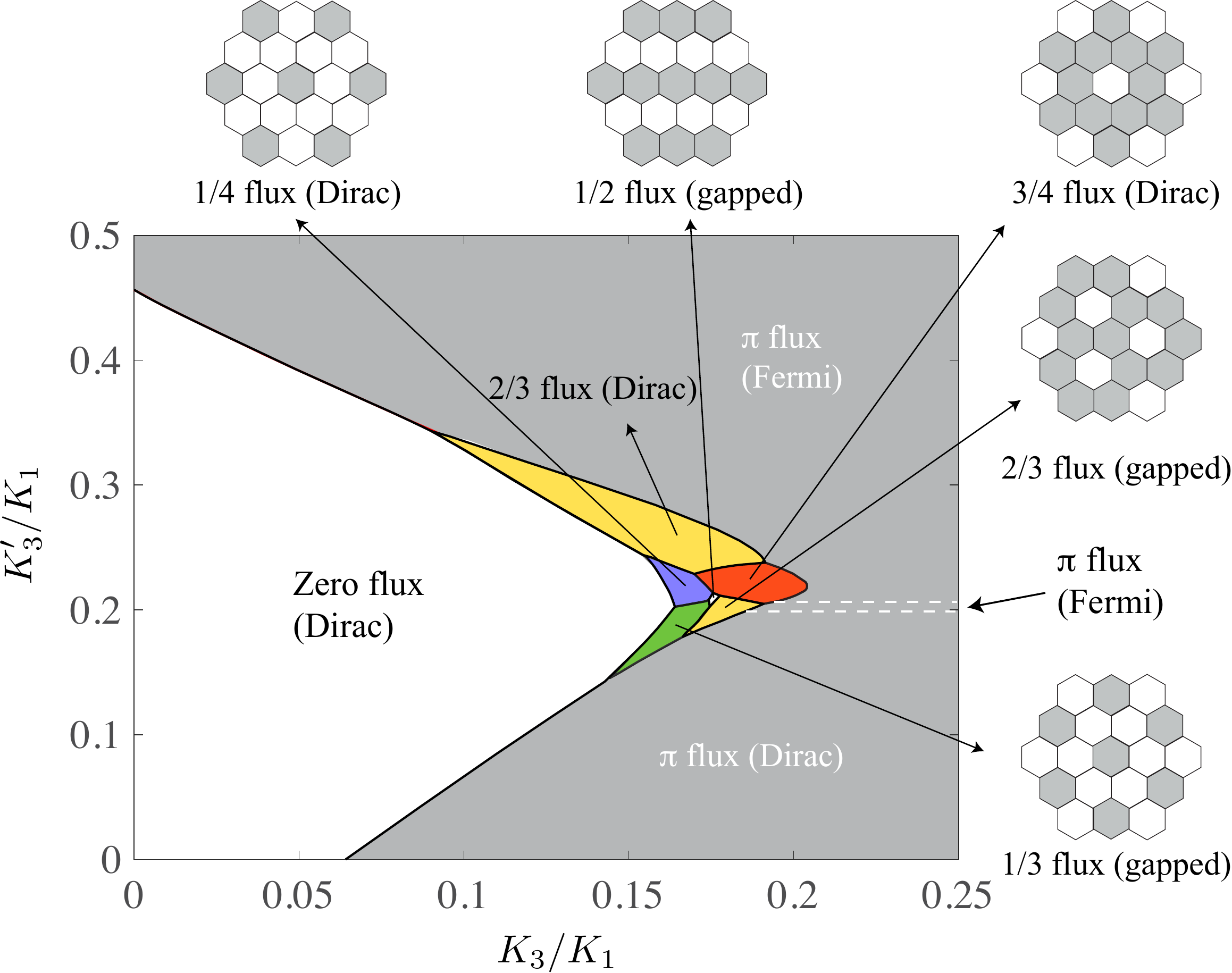}
\caption{Phase diagram of the extended Kitaev model. Flux
configurations of distinct vison crystals (colored phases) are
depicted in separate panels; the presence (absence) of a flux is
marked at each plaquette by gray (white) filling.}
\label{fig:diagram}
\end{figure}

In analogy with how three-spin interactions may be obtained from a
Zeeman field~\cite{kitaev2006anyons}, the four-spin interactions in
Eq.~(\ref{eq:fourspin}) can in principle be generated by a
perturbative treatment of Heisenberg and/or symmetric off-diagonal
($\Gamma$) interactions on top of the pure Kitaev model. Taking a
more universal approach and considering Eq.~(\ref{eq:Hextended}) as
an effective low-energy theory for the Majorana fermions
\cite{song2016low}, we know that generic time-reversal-symmetric
perturbations to ${\cal H}_{K_1}$ must generate all Majorana terms
that are consistent with the projective symmetries of the Kitaev
spin liquid \cite{you2012doping}. Given that all interaction terms
are irrelevant and second-neighbor hopping terms are forbidden by
time reversal, Eq.~(\ref{eq:Hextended}) is the most natural
effective theory beyond the pure Kitaev model.

{\it Phase diagram.} The ground state of ${\cal H}_{K_1}$ belongs to
the zero-flux sector, characterized by $W_p = +1$ for all $p$
\cite{kitaev2006anyons, Lieb94}. In the presence of the additional
interactions, however, the ground state may belong to a wide range
of different flux sectors, as shown by the $T=0$ phase diagram in
Fig.~\ref{fig:diagram}. This phase diagram is obtained from a simple
variational analysis, by comparing the energies of the seven flux
sectors appearing in the diagram on finite lattices of $48 \times
48$ unit cells \footnote{We verified that fluctuations in the phase
boundaries due to finite-size effects become negligibly small for
lattices larger than $36 \times 36$ unit cells.}. Furthermore, it is
fully consistent with unbiased finite-temperature MC simulations,
discussed in a later section \footnote{We verified this statement by
running unbiased MC simulations for multiple randomly chosen points
within each phase on finite lattices of $12 \times 12$ unit cells.}.

We first concentrate on the two fully symmetric non-crystal phases
occupying most of the phase diagram: the zero-flux phase, which has
no fluxes at any plaquettes, and the $\pi$-flux phase, which has a
$\mathbb{Z}_2$ flux at each plaquette. For $K_3 = K_3' = 0$, the
creation of each $\mathbb{Z}_2$ flux with $W_p = -1$ costs a finite
energy $\Delta \approx 0.15 K_1$, and the ground state thus belongs
to the zero-flux sector. For $K_3 / K_1 > 0$, the $K_1$ and $K_3$
terms in Eq.~(\ref{eq:Hextended}) give rise to a frustrated Majorana
hopping and hence an increase in the ground-state energy. However,
due to the two paths between any two opposite sites $i$ and $l$
around a plaquette $p$ [see Fig.~\ref{fig:model}(b)], there are two
equivalent hopping terms $\propto i K_3 c_i c_l$ in
Eq.~(\ref{eq:Hextended}), which interfere constructively for $W_p =
+1$ and destructively for $W_p = -1$. Consequently, as $K_3 / K_1$
is increased, fluxes are effective in relieving frustration from the
Majorana hopping and thus become energetically favorable. Since the
effective interaction between nearby fluxes is attractive for small
$K_3' / K_1$ \cite{kitaev2006anyons}, the corresponding phase
transition between the zero-flux and the $\pi$-flux phases is
strongly first order.

Increasing $K_3' / K_1$, one can modify this interaction and
stabilize various intermediate phases with nontrivial flux
configurations. Indeed, there are five distinct
translation-symmetry-breaking vison-crystal phases in
Fig.~\ref{fig:diagram}, with their ordering wave vectors ${\bm Q}$
corresponding to either the $\mathrm{K}$ point or the $\mathrm{M}$
point(s) of the Brillouin zone (BZ). The two ${\bm Q} = {\bm
Q}_{\mathrm{K}}$ crystals have supercells of three plaquettes,
containing one vison (``$1/3$ flux crystal'') and two visons
(``$2/3$ flux crystal''), respectively. Since there are three
different $\mathrm{M}$ points, ${\bm Q} = {\bm Q}_{\mathrm{M}}$
crystals can exhibit single-${\bm Q}$ or multi-${\bm Q}$ ordering.
The single-${\bm Q}$ crystal is a stripy configuration,
corresponding to a supercell of two plaquettes containing one vison
(``$1/2$ flux crystal''), while the two triple-${\bm Q}$ crystals
have supercells of four plaquettes, containing one vison (``$1/4$
flux crystal'') and three visons (``$3/4$ flux crystal''),
respectively.

\begin{figure}[tbp]
\centering
\includegraphics[width=0.95\columnwidth]{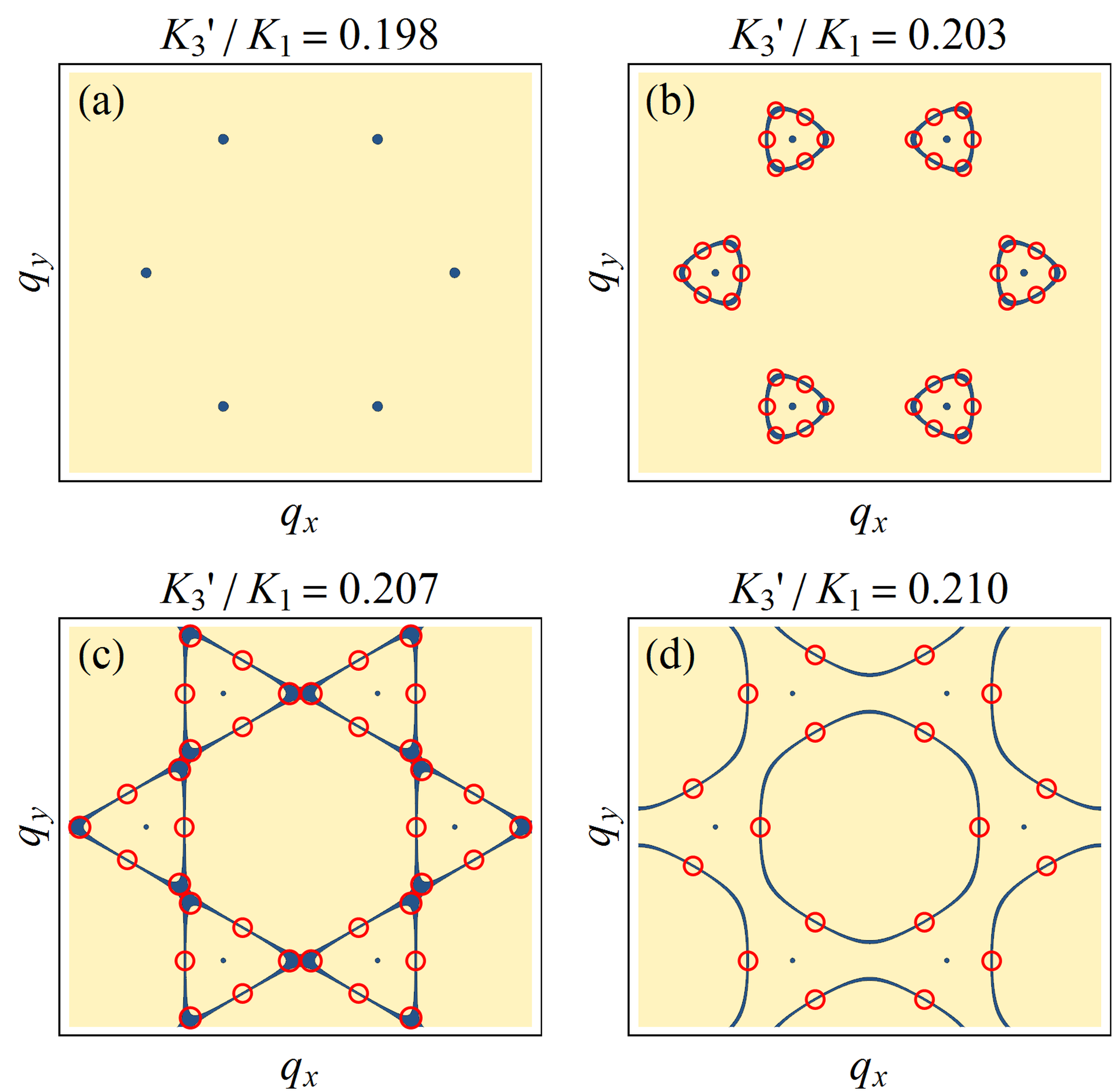}
\caption{Majorana nodal structures (dark blue) in the various
$\pi$-flux phases: the Dirac phase (a), the first Fermi phase (b),
the Lifshitz transition between the two Fermi phases (c), and the
second Fermi phase (d). In the presence of generic further-neighbor
Majorana hopping terms \cite{note0}, each Fermi surface is gapped
out into six Dirac points (red circles).} \label{fig:nodes}
\end{figure}

{\it Majorana problems.} For the different ground-state flux sectors
discussed above, distinct configurations of the gauge fields
$u_{ij}^{\alpha} = \pm 1$ lead to different Majorana Hamiltonians in
Eq.~(\ref{eq:Hextended}). Consequently, each phase in
Fig.~\ref{fig:diagram} has its own Majorana band dispersion and a
corresponding density of states. The low-energy physics, giving rise
to universal signatures in experiments, is determined by the nodal
structures of the Majorana fermions. For the zero-flux phase,
including the pure Kitaev model, as well as for the $1/4$ and $3/4$
flux crystals, the Majorana fermions are gapless at Dirac points and
thus have linear density of states at low energies. For the $1/3$
and $1/2$ flux crystals, the Majorana fermions are fully gapped and
thus have zero density of states below the energy gap. For the $2/3$
flux crystal, there are two disconnected phases where the Majorana
fermions are gapless at Dirac points and fully gapped, respectively
(see Fig.~\ref{fig:diagram}).

Interestingly, the Majorana fermions have more complex nodal
structures in the $\pi$-flux phase. This phase is amenable to a full
analytic understanding as, due to the perfect cancelation of all
$K_3$ terms in Eq.~(\ref{eq:Hextended}), the Majorana problem has
only one dimensionless parameter ratio $\kappa \equiv K_3' / K_1$.
With a simple calculation~\cite{supp}, we find that there are in
fact \emph{three} distinct $\pi$-flux phases characterized by
different Majorana nodal structures.

In particular, there is a $\pi$-flux phase where the Majorana
fermions are gapless at Dirac points only, and another two
$\pi$-flux phases where these Dirac points coexist with Fermi
surfaces (i.e., nodal lines) of distinct topologies (see
Fig.~\ref{fig:nodes}). The dashed lines in Fig.~\ref{fig:diagram}
indicate two subsequent Lifshitz transitions~\cite{Volovik2003}
separating these three phases as a function of increasing $\kappa$.
For $\kappa < 1/5$, the only nodal structures are Dirac points. At
the first Lifshitz transition, $\kappa = 1/5$, small pockets of
Fermi surfaces appear around these Dirac points and gradually expand
as $\kappa$ is further increased. At the second Lifshitz transition,
$\kappa = (\sqrt{2} - 1) / 2 \approx 0.207$, these small pockets
then connect with each other to form larger pockets. We remark that
the Dirac points are located at exactly the same momenta for all
values of $\kappa$.

Such a coexistence of Dirac points and Fermi surfaces is rather
surprising and is not expected to be stable. Instead, due to the
nature of the time-reversal and particle-hole symmetries in the
Majorana problem \cite{hermanns2014quantum}, one would anticipate
only Dirac points to be generically present, as in all the other
phases of Fig.~\ref{fig:diagram}. Indeed, we find that the Fermi
surfaces exist due to the particular simplicity of the problem up to
third-neighbor hopping terms~\cite{supp} and that each Fermi surface
is gapped out into six Dirac points (see Fig.~\ref{fig:nodes}) when
generic fifth-neighbor hopping terms \cite{note0}, respecting the
projective symmetries of the system, are included in
Eq.~(\ref{eq:Hextended}). However, assuming that such terms are
small enough, \emph{approximate} Fermi surfaces are still expected
to be observable in experiments.

{\it Experimental signatures.} The phase diagram in
Fig.~\ref{fig:diagram} contains a rich variety of phases with all
possible Majorana nodal structures in two dimensions, including
Fermi surfaces, Dirac points, and fully gapped scenarios. Due to
their distinct low-energy physics, these phases are characterized by
different experimental signatures. First, we expect the
low-temperature specific heat to behave as $C \propto T$ for Fermi
phases, $C \propto T^2$ for Dirac phases, and $C \propto
e^{-\Delta_{\text{v}} / T}$ for fully gapped phases, where the activated
behavior should be controlled by the vison gap $\Delta_{\text{v}}$ as it is
actually smaller than the Majorana gap. Second, the various Majorana
nodal structures may be distinguished by their low-energy
fingerprints in spectroscopic probes, such as resonant inelastic
x-ray scattering \cite{Gabor2016, Gabor2017}. Third, the Majorana
Fermi surface in the $\pi$-flux phase leads to impurity-induced
Friedel oscillations in the magnetic energy density \cite{supp}. In
turn, such magnetic Friedel oscillations should be measurable with
nuclear magnetic resonance (NMR) as they induce an oscillatory
bond-length modulation via magnetostriction.

\begin{figure}[tbp]
\centering
\includegraphics[width=1.0\columnwidth]{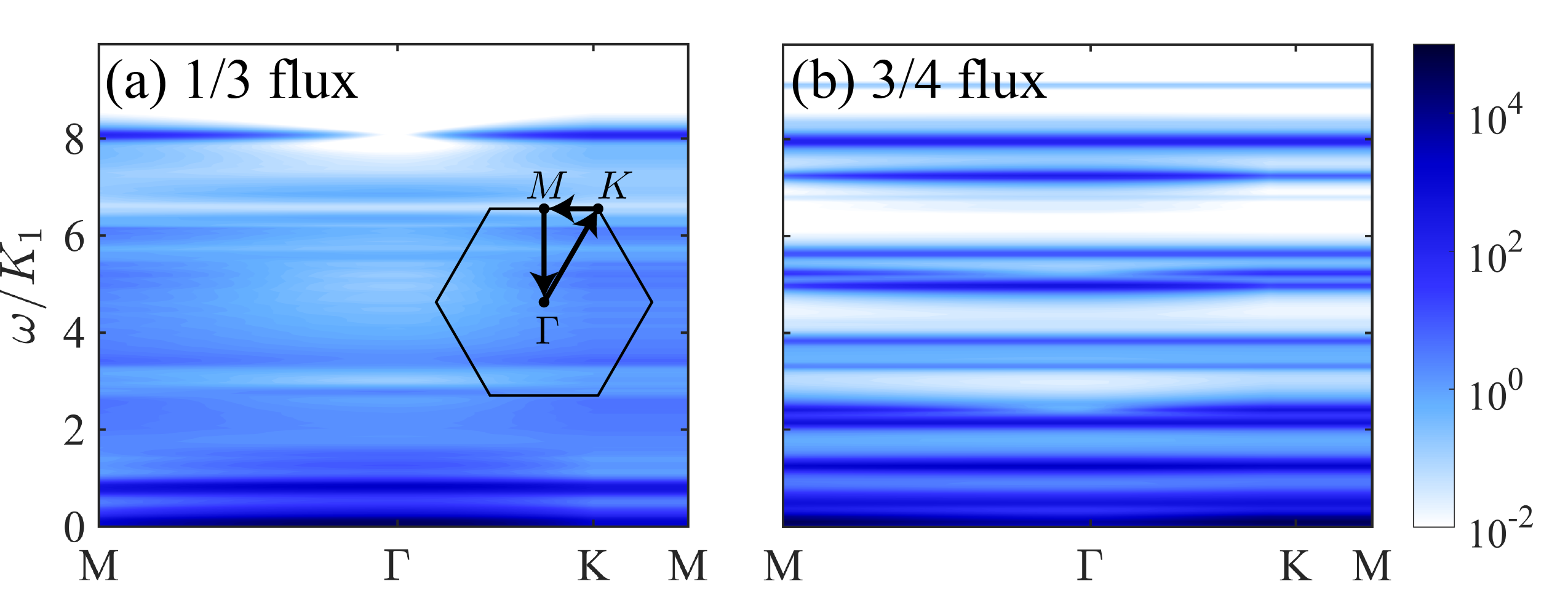}
\caption{Dynamical spin structure factor $S_{zz} ({\bm q}, \omega)$
\cite{supp} for the 1/3 flux crystal (a) and the 3/4 flux crystal
(b) along the path $\mathrm{M}$-$\Gamma$-$\mathrm{K}$-$\mathrm{M}$
in the Brillouin zone [see inset of panel (a)] via the
single-particle approximation of
Ref.~\onlinecite{knolle2015dynamics}.} \label{fig:dynamics}
\end{figure}

For the vison-crystal phases in Fig.~\ref{fig:diagram}, the
spontaneous breaking of translation symmetry leads to further
experimental signatures. First of all, due to magnetostriction, each
vison crystal generates a characteristic bond-length modulation
throughout the lattice, which can be picked up with NMR or elastic
x-ray scattering. Moreover, the enlarged unit cell results in a
larger number of distinct bands for the Majorana fermions and
therefore, in contrast to the pure Kitaev model
\cite{knolle2014dynamics, knolle2015dynamics}, the dynamical spin
structure factor \cite{supp}, directly measurable by inelastic
neutron scattering, has multiple peaks as a function of energy (see
Fig.~\ref{fig:dynamics}). Finally, unlike the fully symmetric
phases, each vison-crystal phase has a finite-temperature phase
transition at a critical temperature $T_c$.

\begin{figure}[tbp]
\centering
\includegraphics[width=0.99\columnwidth]{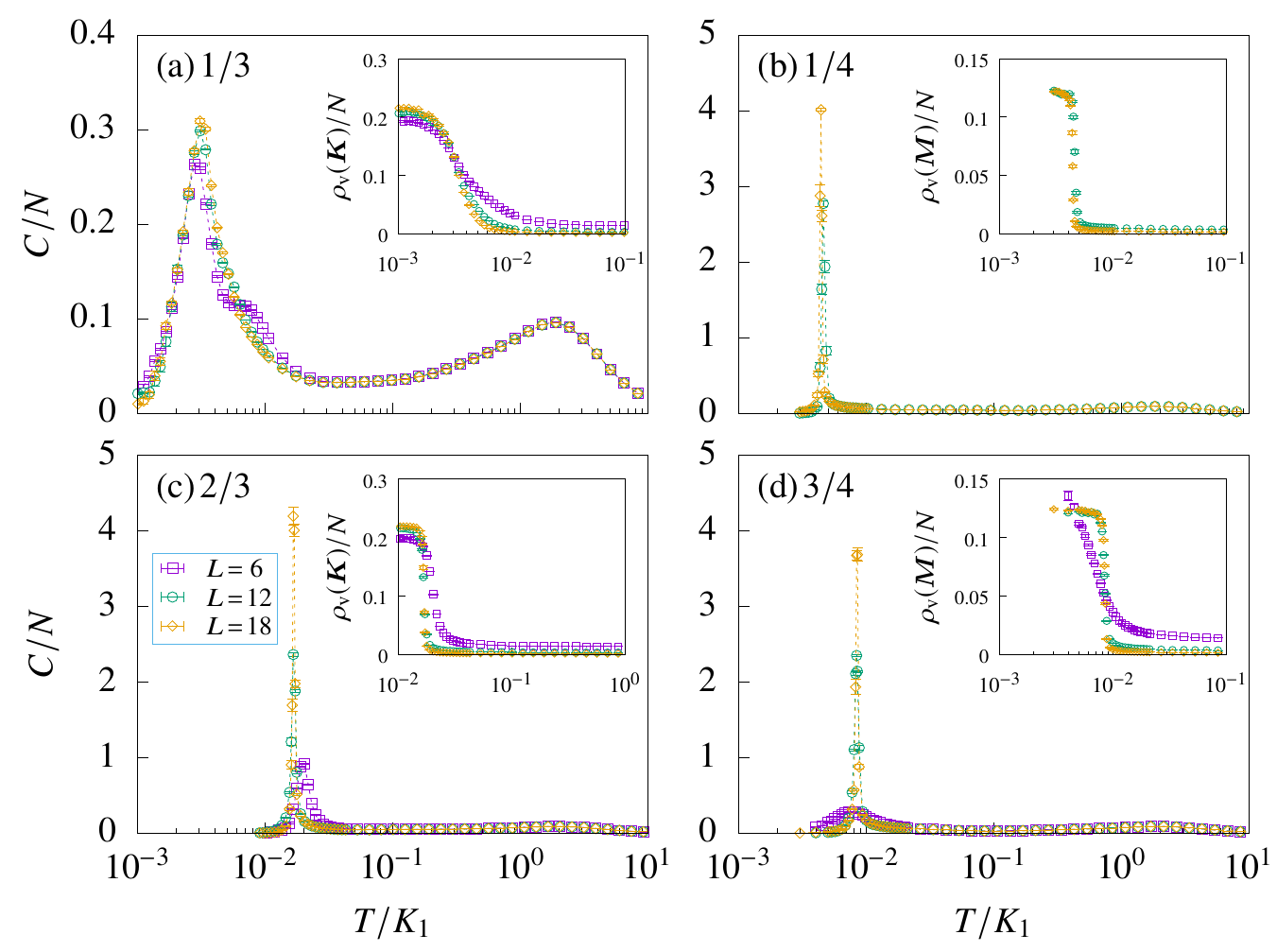}
\caption{Temperature dependence of the specific heat and the
appropriate static vison structure factor for (a) the $1/3$ flux
crystal with $K_3=0.165 K_1$ and $ K_3'=0.19 K_1$, (b) the $1/4$
flux crystal with $K_3=0.165 K_1$ and $K_3'=0.22 K_1$, (c) the $2/3$
flux crystal with $ K_3=0.165 K_1$ and $K_3'=0.26 K_1$, and (d) the
$3/4$ flux crystal with $K_3=0.19 K_1$ and $K_3'=0.22 K_1$ on $L
\times L$ lattices ($L = 6,12,18$) containing $N = 2L^2$ sites.}
\label{fig:thermo}
\end{figure}

{\it Monte Carlo simulations.} To verify the phase diagram in
Fig.~\ref{fig:diagram} and to extract the melting temperatures $T_c$
of the vison crystals, we perform MC simulations of ${\cal H}$ based
on a Metropolis algorithm to update the ``classical'' $\mathbb{Z}_2$
fields $\{ u_{ij}^\alpha = \pm 1 \}$. The energy of each field
configuration is computed by diagonalizing the quadratic Majorana
Hamiltonian in Eq.~\eqref{eq:Hextended} \cite{Nasu2014, Nasu2015} on
$L\times L$ lattices with $L=\{6,12,18\}$ \cite{supp}. For each
temperature, a single run contains 10000 MC sweeps for equilibration
and another 20000 MC sweeps for measurement \footnote{We average
over 8 independent runs to estimate the errors.}.

Figure \ref{fig:thermo} shows our results for the heat capacity
$C(T)$ and the static vison structure factor,
\begin{equation}
\rho_{\text{v}} ({\bm k}) = \frac{1}{L^2} \sum_{p,p'} e^{i {\bm k}
\cdot ({\bm X}_p - {\bm X}_{p'})} \, \langle W_{p} W_{p'} \rangle,
\label{eq:VSF}
\end{equation}
for representative parameters of four different vison crystals,
where ${\bm X}_p$ is the position of plaquette $p$, and ${\bm k}$ is
the ordering wave vector of each vison crystal, corresponding to
either the $\mathrm{K}$ or the $\mathrm{M}$ point of the BZ. We
first observe that, as for the pure Kitaev model, $C(T)$ exhibits
both a high- and a low-temperature peak, which correspond to spinon
and vison excitations, respectively \citep{Nasu2014, Nasu2015}.
However, the low-temperature peak signals the onset of vison-crystal
ordering at $T=T_c$, as confirmed by the sharp growth of the
corresponding Bragg peak in $\rho_{\text{v}} ({\bm k})$. While the
three lattice sizes $L=\{6,12,18\}$ do not facilitate a rigorous
finite-size scaling analysis, the results in Fig.~\ref{fig:thermo}
suggest a first-order crystallization transition for all vison
crystals, except for the $1/3$ flux crystal \footnote{The height of
the peak in $C(T)/L^2$ is proportional to the system volume, $L^2$,
for first-order transitions and to $L^{\alpha/\nu}$ for second-order
transitions.}. Assuming that the transition into the $1/3$ flux
crystal is continuous, it is conjectured to be in the universality
class of the two-dimensional $3$-state Potts model, which in turn
suggests that the height of the peak in $C(T)/L^2$ should be
$\propto L^{\alpha/\nu}$ with critical exponents $\alpha=1/3$,
$\nu=5/6$, and $\alpha/\nu=2/5$ \cite{Nijs1979}. We note that, for
each vison crystal, the critical temperature is $T_c \sim 10^{-2}
K_1$.

{\it Discussion.} By considering a natural extension of the
honeycomb Kitaev model, we have found a rich spectrum of novel
spin-liquid phases that are \emph{not} adiabatically connected to
the original Kitaev model, including a fully symmetric $\pi$-flux
spin liquid, and five distinct symmetry-breaking spin liquids with
various degrees of vison crystallization. In the future, it would be
interesting to study how an external magnetic field affects our spin
liquids. For the Dirac phases, it may generate non-Abelian gapped
spin liquids with distinct Chern numbers of the Majorana fermions
\cite{kitaev2006anyons}. For the gapped phases, it may lead to
nontrivial finite-field phase transitions between topologically
distinct spin liquids.

We thank Arnab Banerjee, Hiroaki Ishizuka, and Johannes Knolle for
useful comments on the manuscript. S.-S.Z., Z.W., and C.D.B.~are
supported by funding from the Lincoln Chair of Excellence in
Physics. The work of G.B.H.~at ORNL was supported by Laboratory
Director's Research and Development funds. This research used
resources of the Oak Ridge Leadership Computing Facility, which is a
DOE Office of Science User Facility supported under Contract
DE-AC05-00OR22725.


%

\clearpage

\begin{widetext}

\setcounter{figure}{0}
\renewcommand{\thefigure}{S\arabic{figure}}
\setcounter{equation}{0}
\renewcommand{\theequation}{S\arabic{equation}}

\subsection{\large Supplemental Material}

\section{Derivation of the Majorana Hamiltonian}

The spin Hamiltonian $\mathcal{H}$ in Eq.~(1) of the main text is
exactly solvable by means of a standard procedure described in
Ref.~1 of the main text. The first step is to introduce four
Majorana fermions $b_j^x$, $b_j^y$, $b_j^z$, and
$c_j^{\phantom{\dag}}$ at each site $j$ of the honeycomb lattice,
and express the spin components $\sigma_j^{x,y,z}$ in terms of these
Majorana fermions as
\begin{equation}
\sigma_j^x = i b_j^x c_j^{\phantom{\dag}}, \qquad \sigma_j^y = i
b_j^y c_j^{\phantom{\dag}}, \qquad \sigma_j^z = i b_j^z
c_j^{\phantom{\dag}}. \label{eq:sigma-1}
\end{equation}
Due to the resulting enlargement of the local Hilbert space, the
four Majorana fermions must be reconciled with the original spin
degree of freedom via the local gauge constraint
\begin{equation}
{-} i \sigma_j^x \sigma_j^y \sigma_j^z = b_j^x b_j^y b_j^z
c_j^{\phantom{\dag}} = 1. \label{eq:gauge}
\end{equation}
In turn, the corresponding gauge redundancy means that the
expressions in Eq.~(\ref{eq:sigma-1}) for the spin components are
not unique; for example, one can use Eq.~(\ref{eq:gauge}) to obtain
the following equivalent expressions:
\begin{equation}
\sigma_j^x = -i b_j^y b_j^z, \qquad \sigma_j^y = -i b_j^z b_j^x,
\qquad \sigma_j^z = -i b_j^x b_j^y. \label{eq:sigma-2}
\end{equation}
Employing Eqs.~(\ref{eq:sigma-1}) and/or (\ref{eq:sigma-2})
appropriately, the two terms ${\cal H}_{K_1}$ and ${\cal H}_{K_3}$
of the spin Hamiltonian $\mathcal{H}$ then become
\begin{eqnarray}
{\cal H}_{K_1} &=& - K_1  \sum_{\langle ij \rangle_{\alpha}}
\sigma^{\alpha}_i \sigma^{\alpha}_j = - K_1  \sum_{\langle ij
\rangle_{\alpha}} \big( i b_i^{\alpha} c_i^{\phantom{\dag}} \big)
\big( i b_j^{\alpha} c_j^{\phantom{\dag}} \big) = i K_1
\sum_{\langle ij \rangle_{\alpha}} \big( i b_i^{\alpha} b_j^{\alpha}
\big) \, c_i^{\phantom{\dag}} c_j^{\phantom{\dag}},
\nonumber \\
{\cal H}_{K_3} &=& K_3 \sum_{\langle ijkl \rangle_{\alpha \beta
\gamma}} \sigma_{i}^{\alpha} \sigma_{j}^{\gamma} \sigma_{k}^{\alpha}
\sigma_{l}^{\gamma} - K_3^{\prime}  \sum_{\langle ijkl
\rangle_{\alpha \beta \alpha}} \sigma_{i}^{\alpha}
\sigma_{j}^{\gamma} \sigma_{k}^{\gamma} \sigma_{l}^{\alpha}
\label{eq:Hterms} \\
&=& K_3 \sum_{\langle ijkl \rangle_{\alpha \beta \gamma}} \big( i
b_i^{\alpha} c_i^{\phantom{\dag}} \big) \big( {-} i b_j^{\alpha}
b_j^{\beta} \big) \big( {-} i b_k^{\beta} b_k^{\gamma} \big) \big( i
b_l^{\gamma} c_l^{\phantom{\dag}} \big) - K_3^{\prime} \sum_{\langle
ijkl \rangle_{\alpha \beta \alpha}} \big( i b_i^{\alpha}
c_i^{\phantom{\dag}} \big) \big( {-} i b_j^{\alpha} b_j^{\beta}
\big) \big( {-} i b_k^{\alpha} b_k^{\beta} \big) \big( i
b_l^{\alpha} c_l^{\phantom{\dag}} \big) \nonumber \\
&=& i K_3 \sum_{\langle ijkl \rangle_{\alpha \beta \gamma}} \big( i
b_i^{\alpha} b_j^{\alpha} \big) \big( i b_k^{\beta} b_j^{\beta}
\big) \big( i b_k^{\gamma} b_l^{\gamma} \big) \,
c_i^{\phantom{\dag}} c_l^{\phantom{\dag}} + i K_3^{\prime}
\sum_{\langle ijkl \rangle_{\alpha \beta \alpha}} \big( i
b_i^{\alpha} b_j^{\alpha} \big) \big( i b_k^{\beta} b_j^{\beta}
\big) \big( i b_k^{\alpha} b_l^{\alpha} \big) \,
c_i^{\phantom{\dag}} c_l^{\phantom{\dag}}. \nonumber
\end{eqnarray}
Substituting Eq.~(\ref{eq:Hterms}) into Eq.~(1) of the main text,
and introducing the $\mathbb{Z}_2$ gauge fields $u_{ij}^{\alpha}
\equiv i b_i^{\alpha} b_j^{\alpha}$, one immediately recovers the
Majorana Hamiltonian in Eq.~(3) of the main text. Since the
$\mathbb{Z}_2$ gauge fields are mutually commuting conserved
quantities, $u_{ij}^{\alpha} = \pm 1$, this Majorana Hamiltonian is
quadratic and hence exactly solvable.

\section{Quadratic Majorana problems}

For each phase in Fig.~2 of the main text, the ground-state flux
configuration can be represented with an appropriate gauge
configuration $u_{ij}^{\alpha} = \pm 1$ (see Fig.~\ref{fig:gauge}).
For all phases other than the zero-flux phase, the effective unit
cell of the Majorana fermions is enlarged with respect to the
honeycomb unit cell as a result of physical symmetry breaking (flux
crystallization) and/or ostensible symmetry breaking (gauge freedom
in representing each $\pi$ flux). We label each site of the
honeycomb lattice as $i \equiv (\Theta, {\bm R}, \lambda)$, where
$\Theta = \{A,B\}$ is a sublattice index, ${\bm R}$ is the lattice
vector of the enlarged unit cell, and $\lambda = 1, \dots, n$
specifies the particular honeycomb unit cell within the enlarged
unit cell. Note that $n=1$ for the zero-flux phase, $n=2$ for the
$\pi$-flux phase, and $n>2$ for the flux-crystal phases.

\begin{figure*}[h]
\centering
\includegraphics[width=0.75\textwidth]{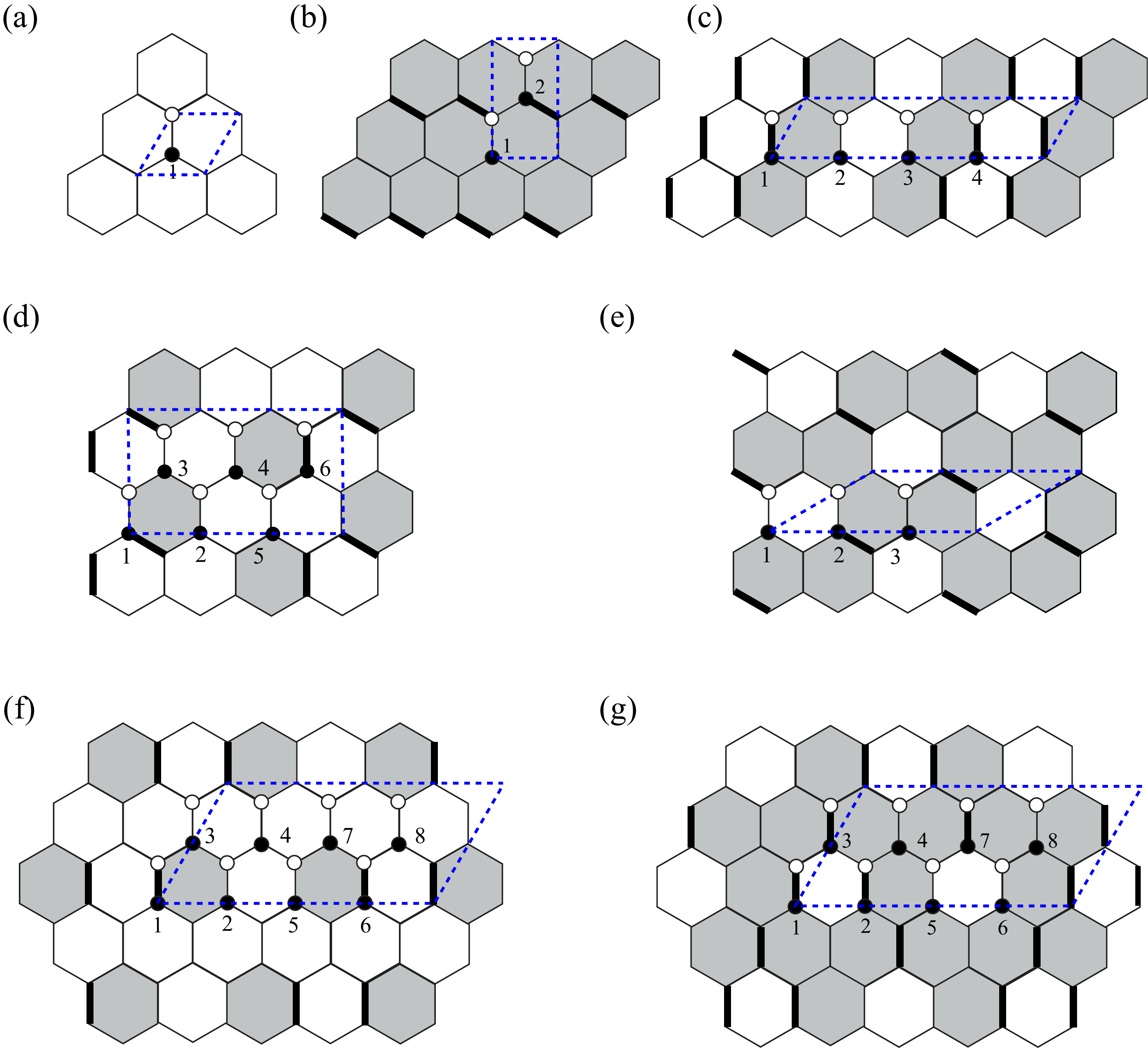}
\caption{Gauge representation of the flux configuration for the
zero-flux phase (a), the $\pi$-flux phase (b), the $1/2$ flux
crystal (c), the $1/3$ flux crystal (d), the $2/3$ flux crystal (e),
the $1/4$ flux crystal (f), and the $3/4$ flux crystal (g). In each
case, bonds with $u_{ij}^{\alpha} = -1$ are denoted by thick lines.
Note that the effective unit cell (blue dashed parallelogram) may
contain several honeycomb unit cells, labeled by $\lambda = 1,
\dots, n$, each containing one $A$ site (black dot) and one $B$ site
(white dot).} \label{fig:gauge}
\end{figure*}

Using this notation, the quadratic Majorana Hamiltonian in Eq.~(3)
of the main text takes the general form
\begin{equation}
\mathcal{H} = \sum_{{\bm R}, {\bm R}'} \sum_{\lambda, \lambda'} i
\mathcal{M}_{{\bm R}' - {\bm R}, \lambda, \lambda'}^{\phantom{\dag}}
\, c_{A, {\bm R}, \lambda}^{\phantom{\dag}} \, c_{B, {\bm R}',
\lambda'}^{\phantom{\dag}}, \label{eq:majorana}
\end{equation}
where each $\mathcal{M}_{{\bm R}' - {\bm R}, \lambda, \lambda'}$ is
a product of gauge fields $u_{ij}^{\alpha} = \pm 1$ along a path
connecting the sites $(A, {\bm R}, \lambda)$ and $(B, {\bm R}',
\lambda')$ occupied by the Majorana fermions $c_{A, {\bm R},
\lambda}^{\phantom{\dag}}$ and $c_{B, {\bm R}',
\lambda'}^{\phantom{\dag}}$. In terms of the momentum-space complex
fermions
\begin{equation}
\psi_{A(B), {\bm q}, \lambda}^{\phantom{\dag}} = \frac{1} {\sqrt{N}}
\sum_{{\bm R}} c_{A(B), {\bm R}, \lambda}^{\phantom{\dag}} \, e^{-i
{\bm q} \cdot {\bm R}}, \qquad \psi_{A(B), {\bm q}, \lambda}^{\dag}
= \psi_{A(B), -{\bm q}, \lambda}^{\phantom{\dag}} = \frac{1}
{\sqrt{N}} \sum_{{\bm R}} c_{A(B), {\bm R},
\lambda}^{\phantom{\dag}} \, e^{i {\bm q} \cdot {\bm R}},
\label{eq:complex}
\end{equation}
where $N$ is the number of sites, the Hamiltonian in
Eq.~(\ref{eq:majorana}) assumes the standard Bogoliubov--de Gennes
form
\begin{eqnarray}
&& \mathcal{H} = 2 \sum_{\pm {\bm q}} \sum_{\lambda, \lambda'}
\left( i \hat{\mathcal{M}}_{{\bm q}, \lambda,
\lambda'}^{\phantom{*}} \, \psi_{A, {\bm q}, \lambda}^{\dag} \,
\psi_{B, {\bm q}, \lambda'}^{\phantom{\dag}} + i
\hat{\mathcal{M}}_{{\bm q}, \lambda, \lambda'}^{*} \, \psi_{A, {\bm
q}, \lambda}^{\phantom{\dag}} \, \psi_{B, {\bm q}, \lambda'}^{\dag}
\right) = 2 \sum_{\pm {\bm q}} \left(
\begin{array}{cc} \psi_{A, {\bm q}}^{\dag} & \psi_{B, {\bm
q}}^{\dag} \end{array} \right) \left(
\begin{array}{cc} 0 & i
\hat{\mathcal{M}}_{{\bm q}}^{\phantom{\dag}} \\ -i
\hat{\mathcal{M}}_{{\bm q}}^{\dag} & 0 \end{array} \right)
\left( \begin{array}{c} \psi_{A, {\bm q}}^{\phantom{\dag}} \\
\psi_{B, {\bm q}}^{\phantom{\dag}} \end{array} \right),
\nonumber \\
&& \qquad \qquad \hat{\mathcal{M}}_{{\bm q}, \lambda,
\lambda'}^{\phantom{\dag}} = \sum_{{\bm r}} \mathcal{M}_{{\bm r},
\lambda, \lambda'}^{\phantom{\dag}} \, e^{i {\bm q} \cdot {\bm r}},
\qquad \qquad \psi_{A(B), {\bm q}}^{\phantom{\dag}} \equiv \left(
\psi_{A(B), {\bm q}, 1}^{\phantom{\dag}}, \dots, \psi_{A(B), {\bm
q}, n}^{\phantom{\dag}} \right)^T, \label{eq:BdG}
\end{eqnarray}
where the summation is over pairs of momenta $\pm {\bm q}$ due to
the particle-hole redundancy $\psi_{A(B), {\bm q}}^{\phantom{\dag}}
= \psi_{A(B), -{\bm q}}^{\dag}$. Using the particular structure of
this quadratic Hamiltonian, reflecting time-reversal symmetry, the
Majorana energies at each momentum $\pm {\bm q}$ are then given by
the singular values of the $n \times n$ matrix
$\hat{\mathcal{M}}_{{\bm q}}^{\phantom{\dag}}$.

\begin{figure}[h]
\includegraphics[width=0.28\textwidth]{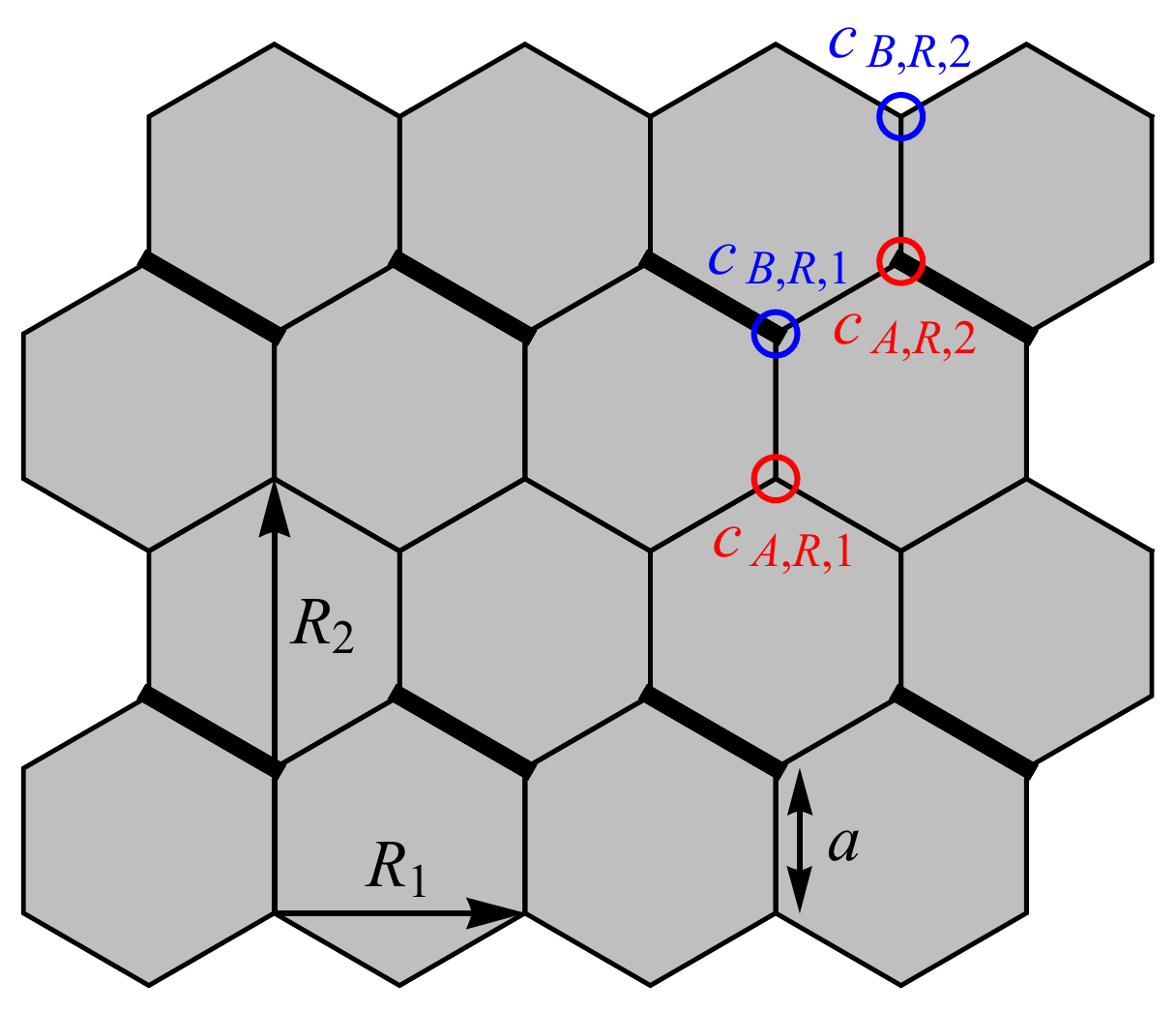}
\caption{Labeling convention for the $\pi$-flux phase, including the
lattice constant $a$, the lattice vectors ${\bm R}_{1,2}$, and the
four sites within each unit cell. The gauge configuration is also
specified; bonds with $u_{ij}^{\alpha} = -1$ are denoted by thick
lines.} \label{fig:gauge-pi}
\end{figure}

\section{Majorana nodal structures}

From Eq.~(\ref{eq:BdG}), the nodal structures of the Majorana
fermions are characterized by vanishing singular values of
$\hat{\mathcal{M}}_{{\bm q}}^{\phantom{\dag}}$ or, equivalently, by
$\det \hat{\mathcal{M}}_{{\bm q}}^{\phantom{\dag}} = 0$. Since $\det
\hat{\mathcal{M}}_{{\bm q}}^{\phantom{\dag}}$ is generically
complex, $\det \hat{\mathcal{M}}_{{\bm q}}^{\phantom{\dag}} = 0$
translates into two independent equations for its real and imaginary
parts. Consequently, for each phase in Fig.~2 of the main text, any
nodal structures are anticipated to be of codimension $2$,
corresponding to point nodes in two dimensions. Indeed, we
generically find that the Majorana fermions are either fully gapped
or gapless at discrete points only.

\begin{figure}[h]
\includegraphics[width=0.5\textwidth]{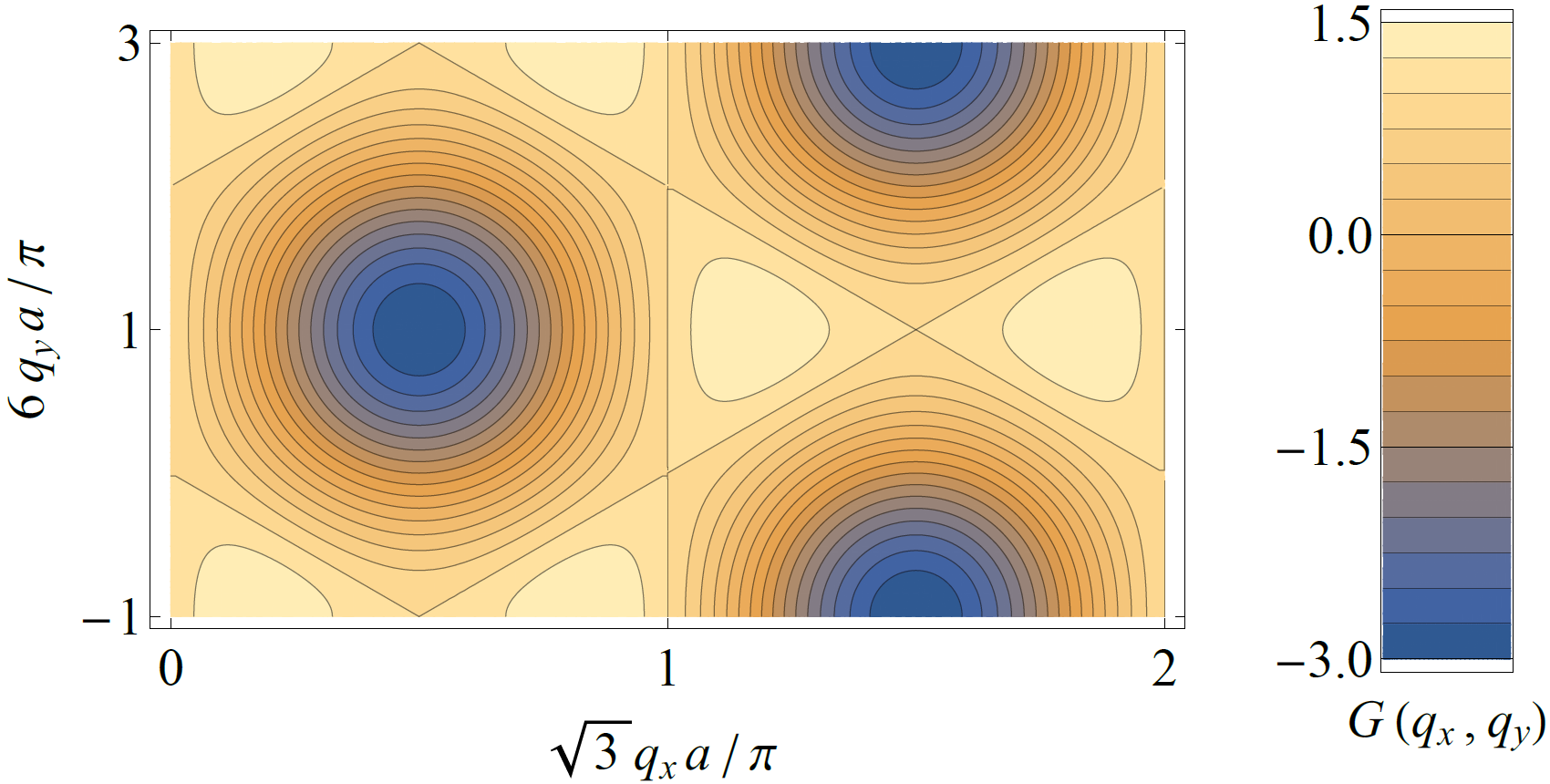}
\caption{Contour plot of the function $G ({\bm q})$ as a function of
the momentum components $q_x$ and $q_y$.} \label{fig:contour}
\end{figure}

\begin{figure}[h]
\includegraphics[width=0.38\textwidth]{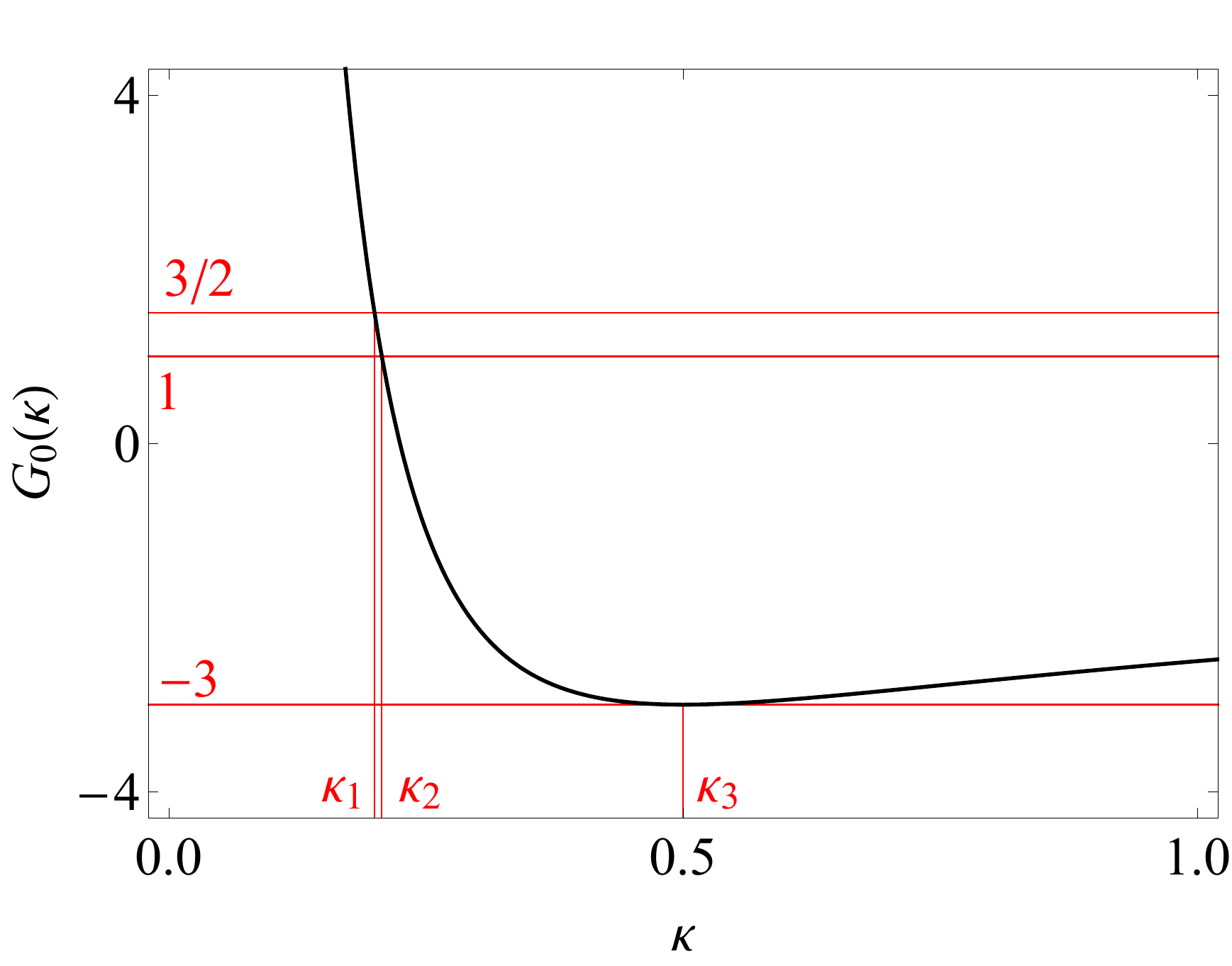}
\caption{Plot of the function $G_0 (\kappa)$ as a function of the
dimensionless parameter $\kappa = K_3' / K_1$.} \label{fig:function}
\end{figure}

However, for the $\pi$-flux phase, if we only consider
first-neighbor and third-neighbor Majorana hopping with amplitudes
$K_1$ and $K_3'$, respectively, the Majorana problem takes a
particularly simple form. Using the labeling convention in
Fig.~\ref{fig:gauge-pi}, the matrix elements of the $2 \times 2$
matrix $\hat{\mathcal{M}}_{{\bm q}}^{\phantom{\dag}}$ in
Eq.~(\ref{eq:BdG}) are given by
\begin{eqnarray}
&& \hat{\mathcal{M}}_{{\bm q}, 1, 1}^{\phantom{\dag}} =
\hat{\mathcal{M}}_{{\bm q}, 2, 2}^{\phantom{\dag}} = K_1 + K_3'
e^{-i {\bm q} \cdot ({\bm R}_1 + {\bm R}_2)} - K_3' e^{i {\bm q}
\cdot ({\bm R}_1 - {\bm R}_2)},
\nonumber \\
&& \hat{\mathcal{M}}_{{\bm q}, 1, 2}^{\phantom{\dag}} = K_1 e^{-i
{\bm q} \cdot ({\bm R}_1 + {\bm R}_2)} + K_1 e^{-i {\bm q} \cdot
{\bm R}_2} + K_3' e^{-i {\bm q} \cdot (2{\bm R}_1 + {\bm R}_2)} -
K_3' e^{-i {\bm q} \cdot {\bm R}_1} + K_3' + K_3' e^{i {\bm q} \cdot
({\bm R}_1 -
{\bm R}_2)}, \qquad \label{eq:pi-M} \\
&& \hat{\mathcal{M}}_{{\bm q}, 2, 1}^{\phantom{\dag}} = K_1 - K_1
e^{i {\bm q} \cdot {\bm R}_1} - K_3' e^{-i {\bm q} \cdot {\bm R}_1}
+ K_3' e^{i {\bm q} \cdot {\bm R}_2} + K_3' e^{i {\bm q} \cdot ({\bm
R}_1 + {\bm R}_2)} + K_3' e^{2i {\bm q} \cdot {\bm R}_1}, \nonumber
\end{eqnarray}
and the determinant of $\hat{\mathcal{M}}_{{\bm
q}}^{\phantom{\dag}}$ readily factorizes into the product form
\begin{equation}
\det \hat{\mathcal{M}}_{{\bm q}}^{\phantom{\dag}} = K_1^2 F ({\bm
q}) \left[ 1 - 2 \kappa (2 + \kappa) - 2 \kappa^2 G ({\bm q})
\right], \label{eq:pi-det}
\end{equation}
where $\kappa \equiv K_3' / K_1$ is the dimensionless third-neighbor
hopping amplitude, and the two functions $F ({\bm q})$ and $G ({\bm
q})$ are
\begin{eqnarray}
F ({\bm q}) &=& 1 + 2i e^{-3i q_y a } \sin \left( \sqrt{3} q_x a
\right), \nonumber \\
G ({\bm q}) &=& \cos \left( 2 \sqrt{3} q_x a \right) - 2 \sin \left(
2 \sqrt{3} q_x  a \right) \sin \left( 3 q_y a \right).
\label{eq:pi-FG}
\end{eqnarray}
While the function $F ({\bm q})$ is complex, and the solutions of $F
({\bm q}) = 0$ thus give point nodes at $q_x a = \pm \pi / (6
\sqrt{3}) + 2 \pi n_x / \sqrt{3}$ and $q_y a = \mp \pi / 6 + 2 \pi
n_y / 3$ as well as at $q_x a = \pm 5 \pi / (6 \sqrt{3}) + 2 \pi n_x
/ \sqrt{3}$ and $q_y a = \mp \pi / 6 + 2 \pi n_y / 3$ (with $n_x,
n_y \in \mathbb{Z}$), the function $G ({\bm q})$ plotted in
Fig.~\ref{fig:contour} is real, with a minimum value
$G_{\mathrm{min}} = -3$, a maximum value $G_{\mathrm{max}} = 3/2$,
and a critical value $G_{\mathrm{vH}} = 1$ corresponding to a van
Hove singularity. Consequently, the solutions of $1 - 2 \kappa (2 +
\kappa) - 2 \kappa^2 G ({\bm q}) = 0$ generically correspond to
nodal lines along the contours of Fig.~\ref{fig:contour} given by
\begin{equation}
G ({\bm q}) = G_0 (\kappa) \equiv \frac{1 - 2 \kappa (2 + \kappa)}
{2 \kappa^2} \, . \label{eq:pi-G}
\end{equation}
To analyze these nodal lines, we plot $G_0 (\kappa)$ in
Fig.~\ref{fig:function} and find three critical values of $\kappa$
between $0$ and $1$:
\begin{equation}
\kappa_1 = 0.2, \qquad \kappa_2 = \frac{1}{2} \big( \sqrt{2} - 1
\big) \approx 0.207, \qquad \kappa_3  = 0.5. \label{eq:pi-kappa}
\end{equation}
For $\kappa < \kappa_1$, we obtain $G_0 (\kappa) >
G_{\mathrm{max}}$, and Eq.~(\ref{eq:pi-G}) has no solutions. For
$\kappa_1 < \kappa < \kappa_2$, we obtain $G_{\mathrm{vH}} < G_0
(\kappa) < G_{\mathrm{max}}$, and the solutions of
Eq.~(\ref{eq:pi-G}) are nodal lines surrounding the maxima of $G
({\bm q})$. Interestingly, these maxima coincide with the point
nodes characterized by $F ({\bm q}) = 0$. Finally, for $\kappa >
\kappa_2$, we obtain $G_{\mathrm{min}} \leq G_0 (\kappa) <
G_{\mathrm{vH}}$, and the solutions of Eq.~(\ref{eq:pi-G}) are nodal
lines surrounding the minima of $G ({\bm q})$. In particular, for
$\kappa = \kappa_3$, these nodal lines contract to point nodes as
$G_0 (\kappa) = G_{\mathrm{min}}$.

\section{Magnetic Friedel oscillations}

In principle, a non-magnetic impurity, such as a spin vacancy [see
Fig.~\ref{fig:friedel}(a)], can be used as a ``physical probe'' to
distinguish between the $\pi$-flux phases with Dirac points and
Fermi surfaces of Majorana fermions. Such an impurity induces a
local modulation of the bond energy $E_{\langle ij \rangle_{\alpha}}
\propto \langle \sigma_i^{\alpha} \sigma_j^{\alpha} \rangle$, which
decays with a power law as a function of distance; Friedel
oscillations are expected to be present (absent) in this decay if
the Majorana fermions are gapless at Fermi surfaces (Dirac points).
We therefore calculate the radial Fourier transform of the
bond-energy modulation,
\begin{equation}
\Delta E(p) = \frac{1}{N} \sum_{\langle ij \rangle_{\alpha}}
e^{-ipr_{\langle ij \rangle_{\alpha}}} \left[ E_{\langle ij
\rangle_{\alpha}}^{\phantom{(0)}} - E_{\langle ij
\rangle_{\alpha}}^{(0)} \right], \label{eq:friedel}
\end{equation}
in both phases [see Fig.~\ref{fig:friedel}(b)], where $r_{\langle ij
\rangle_{\alpha}}$ is the distance of the bond $\langle ij
\rangle_{\alpha}$ from the impurity, and $E_{\langle
ij\rangle_{\alpha}}^{(0)}$ is the bond energy in the absence of the
impurity. As expected, in the Dirac phase, $\Delta E(p)$ is peaked
at $p=0$, while in the Fermi phase, its peak is shifted to $p
\approx 2q_F$, where $q_F$ is the radius of the Fermi surface [see
Fig.~\ref{fig:friedel}(c)].

\begin{figure}[h]
\centering
\includegraphics[width=0.8\textwidth]{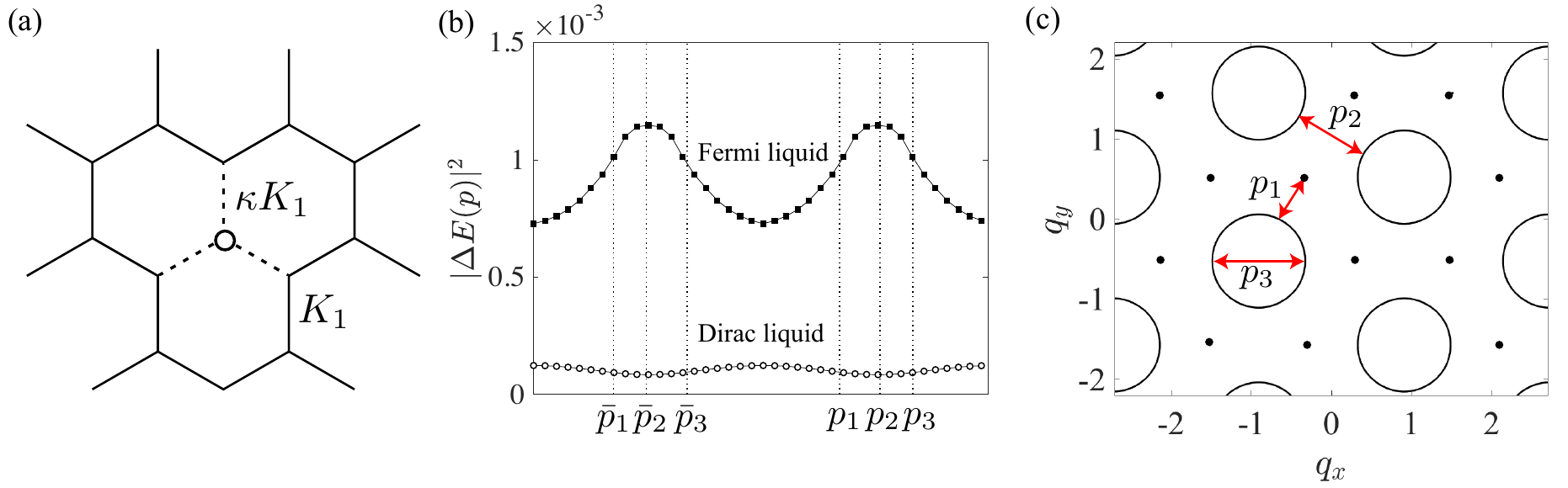}
\caption{(a) Local distortion of the Kitaev interactions ($\kappa
\neq 1$) around a nonmagnetic impurity; the particular case of a
spin vacancy corresponds to $\kappa = 0$. (b) Radial Fourier
transform of the bond-energy modulation around the impurity for the
$\pi$-flux phases with Dirac nodes (white circles) and Fermi
surfaces (black squares) of Majorana fermions. (c) For the Fermi
phase, the peak wave vector of Friedel oscillations is compared with
the characteristic dimensions of the Fermi surfaces.}
\label{fig:friedel}
\end{figure}

\section{Dynamical spin structure factor}

The dynamical spin structure factor $S_{\mu \nu} ({\bm q},\omega)$
is probed experimentally by inelastic neutron scattering. At $T=0$,
it is given by the spatial and temporal Fourier transform of the
spin-spin correlation function in the ground state:
\begin{equation}
S_{\mu \nu} ({\bm q}, \omega) = \frac{1}{2\pi N} \sum_{i,j}
\int_{-\infty}^{+\infty} dt \, e^{i \omega t - i{\bm q}\cdot ({\bm
r}_j - {\bm r}_i)} \, \big\langle \sigma_i^{\mu} (t) \sigma_j^{\nu}
(0) \big\rangle, \label{eq:Sqw}
\end{equation}
where $\sigma_i^{\mu} (t) \equiv e^{i \mathcal{H} t} \sigma_i^{\mu}
e^{-i \mathcal{H} t}$, and ${\bm r}_i$ is the position of site $i$.
For the general Hamiltonian $\mathcal{H}$ in Eq.~(1) of the main
text, the spin-spin correlation function $\langle \sigma_i^{\mu} (t)
\sigma_j^{\nu} (0) \rangle$ vanishes unless $\mu = \nu$. Moreover,
its has an extremely limited range: $\langle \sigma_i^{\mu} (t)
\sigma_j^{\mu} (0) \rangle$ is only nonzero if $i$ and $j$ are the
same site ($i = j$) or if they are nearest-neighbor sites connected
by a $\mu$ bond $\langle ij \rangle_{\mu}$. The structure factor in
Eq.~(\ref{eq:Sqw}) is thus generally given by
\begin{equation}
S_{\mu \nu} ({\bm q}, \omega) = \delta_{\mu \nu} \left[ S_{\mu
\mu}^{(0)} (\omega) + 2 \cos({\bm q} \cdot {\bm n}_{\mu}) \, S_{\mu
\mu}^{(1)} (\omega) \right], \label{eq:Sqw-mod}
\end{equation}
where $S_{\mu \mu}^{(0)} (\omega)$ and $S_{\mu \mu}^{(1)} (\omega)$
are on-site and nearest-neighbor contributions,
\begin{eqnarray}
&& S_{\mu \mu}^{(0)} (\omega) = \frac{1}{N} \sum_i S_{\mu \mu,
i}^{(0)} (\omega), \qquad \qquad S_{\mu \mu, i}^{(0)} (\omega) =
\frac{1}{2\pi} \int_{-\infty}^{+\infty} dt \, e^{i \omega t} \,
\big\langle \sigma_i^{\mu} (t) \sigma_i^{\mu} (0) \big\rangle,
\label{eq:S01} \\
&& S_{\mu \mu}^{(1)} (\omega) = \frac{1}{N} \sum_{\langle ij
\rangle_{\mu}} S_{\mu \mu, \langle ij \rangle_{\mu}}^{(1)} (\omega),
\qquad \, S_{\mu \mu, \langle ij \rangle_{\mu}}^{(1)} (\omega) =
\frac{1}{2\pi} \int_{-\infty}^{+\infty} dt \, e^{i \omega t} \,
\big\langle \sigma_i^{\mu} (t) \sigma_j^{\mu} (0) \big\rangle,
\nonumber
\end{eqnarray}
and ${\bm n}_{\mu}$ is the vector connecting the two sites $i$ and
$j$ along any $\mu$ bond $\langle ij \rangle_{\mu}$. Note that
$\langle \sigma_i^{\mu} (t) \sigma_j^{\mu} (0) \rangle = \langle
\sigma_j^{\mu} (t) \sigma_i^{\mu} (0) \rangle$ and, consequently,
$S_{\mu \mu, \langle ij \rangle_{\mu}}^{(1)} (\omega)$ is real due
to time-reversal symmetry.

The general results in Eqs.~(\ref{eq:Sqw-mod}) and (\ref{eq:S01})
are simplified by the unbroken space-group symmetries in each phase
of Fig.~2 in the main text. First of all, due to translation
symmetry, there are only a finite number of inequivalent sites $i$
and inequivalent $\mu$ bonds $\langle ij \rangle_{\mu}$, and the
infinite averages in Eq.~(\ref{eq:S01}) can thus be substituted with
finite averages over the $n_{(0)}$ inequivalent sites and the
$n_{(1)}$ inequivalent $\mu$ bonds. In particular, $n_{(0)} =
n_{(1)} = 1$ for each symmetric phase, while $n_{(0)} = n_{(1)} = r$
for each flux crystal with a supercell of $r$ plaquettes. Moreover,
for all phases other than the $1/2$ flux crystal, there is a
threefold rotation symmetry around the center of some plaquette,
which permutes the spin components as $z \rightarrow x \rightarrow y
\rightarrow z$ and therefore implies $S_{xx}^{(0,1)} (\omega) =
S_{yy}^{(0,1)} (\omega) = S_{zz}^{(0,1)} (\omega)$. For the $1/2$
flux crystal, this rotation symmetry is spontaneously broken; if the
flux stripes are perpendicular to the $y$ bonds, as in
Fig.~\ref{fig:gauge}(c), the remaining point-group symmetries still
imply $S_{xx}^{(0,1)} (\omega) = S_{zz}^{(0,1)} (\omega)$.

\begin{figure*}[h]
\centering
\includegraphics[width=0.9\textwidth]{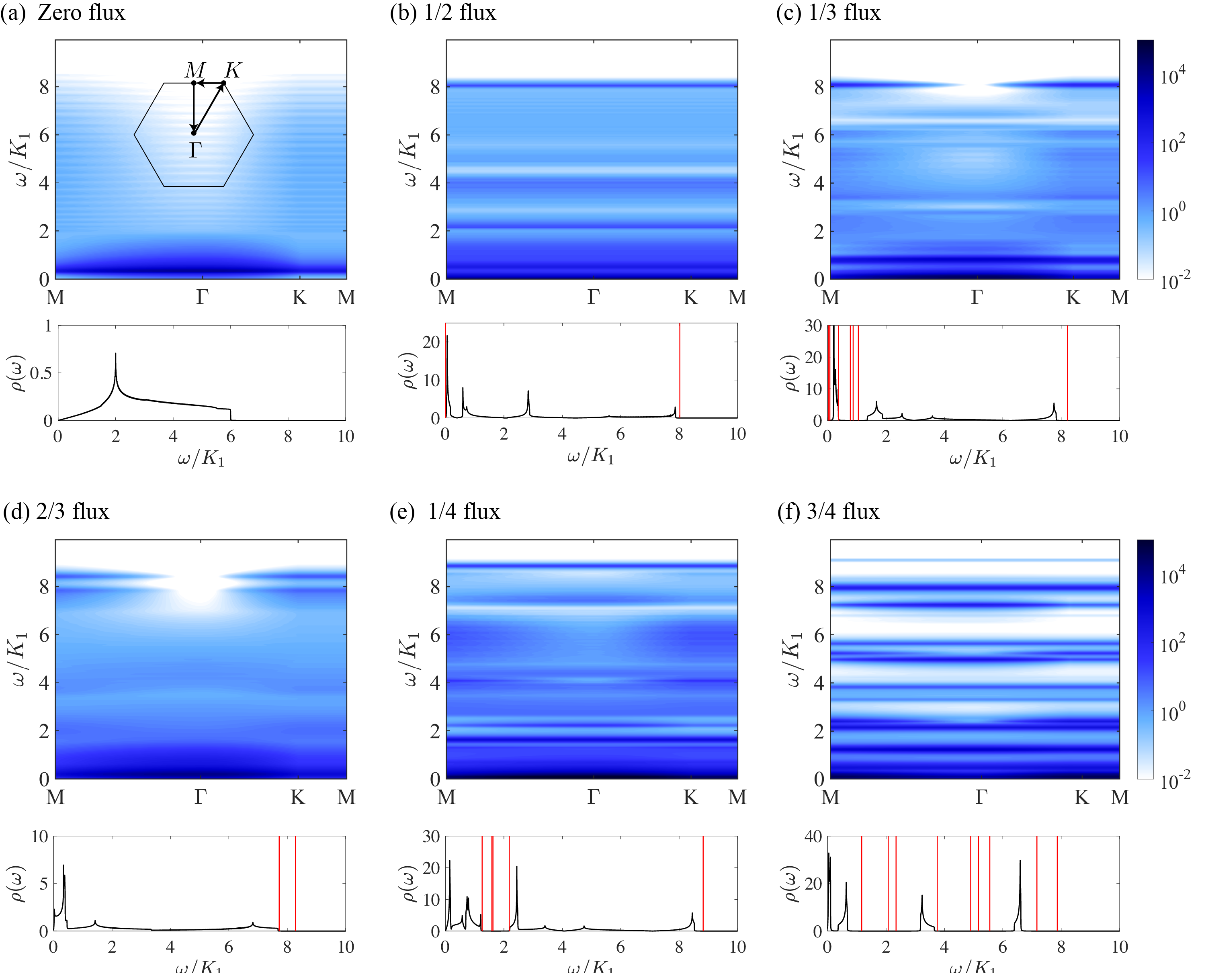}
\caption{Upper panel of (a)-(f): Dynamical spin structure factor
$S_{zz} ({\bm q}, \omega)$ at $T=0$ along the path
$\mathrm{M}$-$\Gamma$-$\mathrm{K}$-$\mathrm{M}$ [see inset of (a)]
for various phases in Fig.~2 of the main text. Each response is
convolved, as a function of energy, with a Lorentzian broadening
function of width $\eta = 0.05 K_1$. Lower panel of (a)-(f):
Majorana density of states $\rho (\omega)$ for each phase. Red
vertical lines are $\delta$ peaks corresponding to additional
localized states that appear in the intermediate flux sector of the
Lehmann representation.} \label{fig:dynamics-sm}
\end{figure*}

In Fig.~\ref{fig:dynamics-sm}, we present the dynamical spin
structure factor $S_{zz} ({\bm q}, \omega)$ at $T=0$ for each
flux-crystal phase along the high-symmetry path
$\mathrm{M}$-$\Gamma$-$\mathrm{K}$-$\mathrm{M}$ in the Brillouin
zone [see inset of panel (a)]. Following the few-particle approach
in Ref.~\onlinecite{knolle2015dynamics}, we take the Lehmann
representation of $S_{zz} (\bm q, \omega)$ and restrict our
attention to intermediate states containing a single Majorana
excitation. If this approximation is valid, the calculated response
should approximately satisfy the sum rule $\int d\omega \,
S_{zz}^{(0)} (\omega) = 1$; for all of the results in
Fig.~\ref{fig:dynamics-sm}, we find that $\int d\omega \,
S_{zz}^{(0)} (\omega) > 0.6$.

The energy dependence of the dynamical spin structure factor
reflects the Majorana density of states in the intermediate flux
sector of the Lehmann representation (see
Fig.~\ref{fig:dynamics-sm}). Due to the two fluxes created (or
destroyed) by the spin operator $\sigma_l^z$, the Majorana fermions
in the intermediate flux sector are perturbed with respect to the
ground-state flux sector, and they may even form localized states
around the site $l$. Such a localized state corresponds to a delta
peak in the density of states and thus gives rise to a sharp feature
in $S_{zz} ({\bm q}, \omega)$. Physically, it can be understood as a
magnon bound state, $\sigma_l^z = i b_l^z c_l^{\phantom{\dag}}$, of
a bond Majorana fermion $b_l^z$ and a matter Majorana fermion
$c_l^{\phantom{\dag}}$.

\section{Monte Carlo implementation}

The Monte Carlo (MC) simulations are implemented on an $L \times L$
honeycomb lattice with dimensions $\bm{L}_{1} = L \bm{a}_{1}$ and
$\bm{L}_{2} = L \bm{a}_{2}$ and periodic boundary conditions (PBC)
in both directions (see Fig.~\ref{fig:MC}). However, using these
boundary conditions, a naive numerical implementation would be
extremely tedious, and we therefore simplify the procedure by
removing a single bond from the lattice.

\begin{figure}
\centering
\includegraphics[width=0.5\columnwidth]{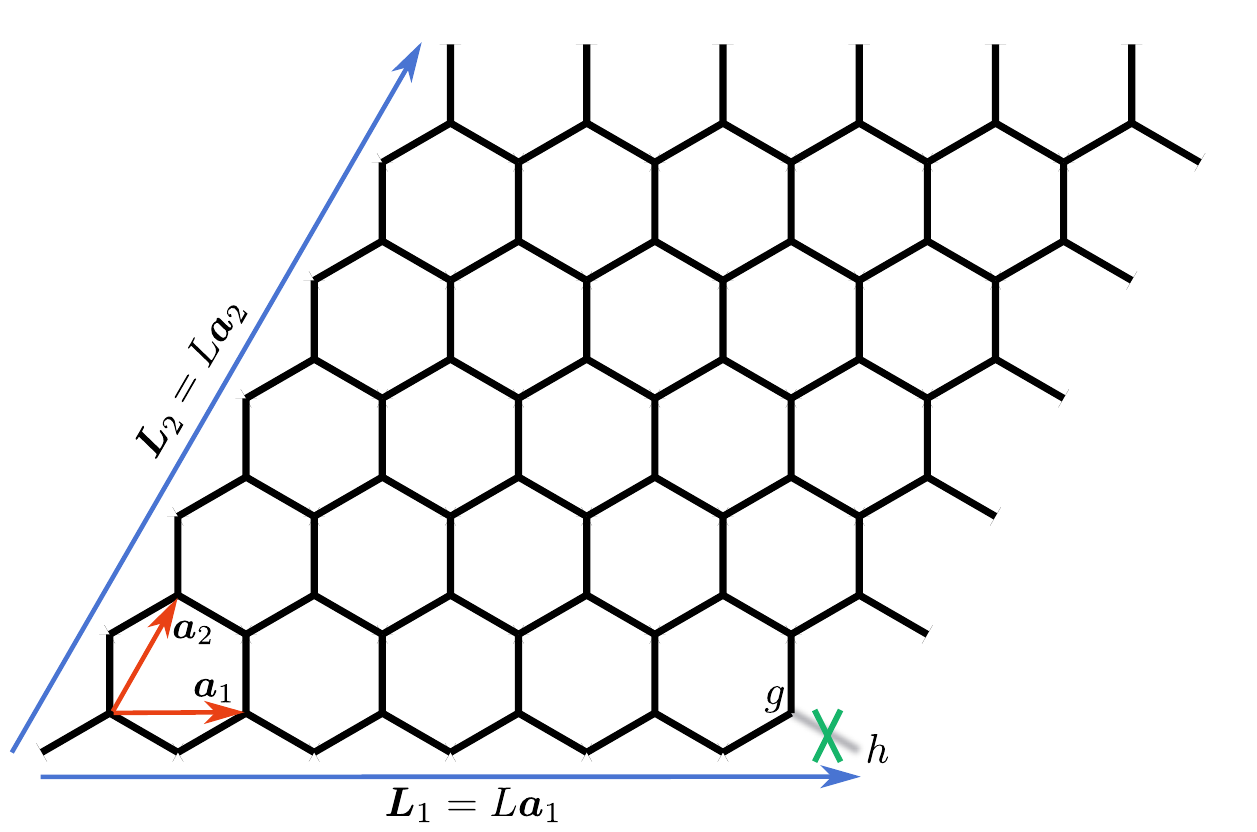}
\caption{Honeycomb lattice for the Monte Carlo simulations. We
employ periodic boundary conditions in both directions but remove a
single $y$ bond (marked by the cross) to avoid a tedious numerical
procedure.} \label{fig:MC}
\end{figure}

Indeed, since the Hilbert space is enlarged by the introduction of
the Majorana fermions, any state in the Majorana representation must
be projected back into the physical Hilbert space. Due to this
projection, distinct states in the Majorana representation may
correspond to the same physical state, and certain states in the
Majorana representation may not correspond to any physical state at
all. In fact, each physical state can be represented by $2^{N-1}$
distinct Majorana states, and the total fermion number, $N_b + N_c$,
including bond and matter fermions, is even for all of them:
\begin{equation}
(-1)^{N_b} (-1)^{N_c} = \Bigg[ \prod_{\langle jk \rangle_{\alpha}}
u_{jk}^{\alpha} \Bigg] \Bigg[ \prod_{\langle jk \rangle_z} w_{jk}
\Bigg] = +1, \label{eq:parity}
\end{equation}
where $u_{jk}^{\alpha} \equiv i b_j^{\alpha} b_k^{\alpha}$ (as in
the main text) and $w_{jk} \equiv i c_j c_k$. The remaining Majorana
states with odd fermion number do not correspond to any physical
state as they are annihilated by the projection. The partition
function is then
\begin{equation}
Z = \frac{1} {2^{N-1}} \,
\text{Tr}_{\{u_{jk}^{\alpha}\}_{\text{even}}}
\text{Tr}_{\{w_{jk}\}_{\text{even}}} e^{-\beta \mathcal{H}} +
\frac{1}{2^{N-1}} \, \text{Tr}_{\{u_{jk}^{\alpha}\}_{\text{odd}}}
\text{Tr}_{\{w_{jk}\}_{\text{odd}}} e^{-\beta \mathcal{H}},
\label{eq:Z-1}
\end{equation}
where $\text{Tr}_{\{u_{jk}^{\alpha}\}_{\text{even}}}$
($\text{Tr}_{\{u_{jk}^{\alpha}\}_{\text{odd}}}$) sums over all
bond-fermion configurations $\{u_{jk}^{\alpha}\}$ with
$\prod_{\langle jk \rangle_{\alpha}} u_{jk}^{\alpha} = +1$ ($-1$),
while $\text{Tr}_{\{w_{jk}\}_{\text{even}}}$
($\text{Tr}_{\{w_{jk}\}_{\text{odd}}}$) sums over all matter-fermion
configurations $\{w_{jk}\}$ with $\prod_{\langle jk \rangle_z}
w_{jk} = +1$ ($-1$).

For the relatively small system sizes accessible with MC, it is
important to take proper care of the even/odd projections, which in
turn makes the MC implementation extremely tedious. However, the
computation is simplified tremendously if we remove a single bond
$\langle gh \rangle_y$ from the lattice (see Fig.~\ref{fig:MC}) by
switching off all interactions that involve both sites $g$ and $h$
\cite{Nasu2014}. In this case, the Hamiltonian $\mathcal{H}$ does
not depend on $u_{gh}^y$ and, by switching between $u_{gh}^y = \pm
1$, one can switch the bond-fermion parity $\prod_{\langle jk
\rangle_{\alpha}} u_{jk}^{\alpha}$ without affecting the spectrum of
the matter fermions at all. The partition function in
Eq.~(\ref{eq:Z-1}) can then be written as
\begin{equation}
Z = \frac{1} {2^{N-1}} \, \text{Tr}_{\{u_{jk}^{\alpha}\}'} \Big[
\text{Tr}_{\{w_{jk}\}_{\text{even}}} e^{-\beta \mathcal{H}} +
\text{Tr}_{\{w_{jk}\}_{\text{odd}}} e^{-\beta \mathcal{H}} \Big] =
\frac{1} {2^{N-1}} \, \text{Tr}_{\{u_{jk}^{\alpha}\}'}
\text{Tr}_{\{w_{jk}\}} e^{-\beta \mathcal{H}}, \label{eq:Z-2}
\end{equation}
where $\text{Tr}_{\{u_{jk}^{\alpha}\}'}$ sums over all
configurations of those bond fermions that do not correspond to the
removed bond $\langle gh \rangle_y$, and $\text{Tr}_{\{w_{jk}\}}$
sums over all configurations of the matter fermions. Finally, the
partition function takes the form
\begin{equation}
Z = \frac{1} {2^{N-1}} \, \text{Tr}_{\{u_{jk}^{\alpha}\}'} \prod_n
\left( e^{\beta \epsilon_n [\{u_{jk}^\alpha\}'] / 2} + e^{-\beta
\epsilon_n [\{u_{jk}^\alpha\}'] / 2} \right) \equiv \frac{1}
{2^{N-1}} \, \text{Tr}_{\{u_{jk}^{\alpha}\}'} e^{-\beta F_w
[\{u_{jk}^\alpha\}']} \equiv \frac{Z'} {2^{N-1}} \, , \label{eq:Z-3}
\end{equation}
where $\epsilon_n [\{u_{jk}^\alpha\}']$ are the non-negative
single-particle energies of the matter fermions for a given
configuration $\{u_{jk}^{\alpha}\}'$ of the bond fermions. From this
simplified partition function, all other thermodynamic quantities
can then be obtained as described in Ref.~\onlinecite{Nasu2014}. In
particular, the internal energy and the heat capacity are given by
\begin{eqnarray}
&& U = \left\langle E_w [\{u_{jk}^\alpha\}'] \right\rangle, \qquad
E_w [\{u_{jk}^\alpha\}'] \equiv -\sum_n \frac{\epsilon_n
[\{u_{jk}^\alpha\}']} {2} \tanh \frac{\beta \epsilon_n
[\{u_{jk}^\alpha\}']} {2} \, ,
\nonumber \\
&& C = \beta^2 \left[ \left\langle E_w [\{u_{jk}^\alpha\}']^2 -
\frac{\partial E_w [\{u_{jk}^\alpha\}']} {\partial \beta}
\right\rangle - \left\langle E_w [\{u_{jk}^\alpha\}']
\right\rangle^2 \right], \label{eq:UC}
\end{eqnarray}
where $\langle \cdots \rangle$ denotes the thermal average over all
bond-fermion configurations $\{u_{jk}^{\alpha}\}'$:
\begin{equation}
\left\langle O_w [\{u_{jk}^\alpha\}'] \right\rangle \equiv
\frac{1}{Z'} \, \text{Tr}_{\{u_{jk}^{\alpha}\}'} \Big[ O_w
[\{u_{jk}^\alpha\}'] \, e^{-\beta F_w [\{u_{jk}^\alpha\}']} \Big].
\label{eq:average}
\end{equation}
Note that the same simplification in the partition function could
also be achieved by removing several bonds from the lattice; we
remove only one bond to minimize any accompanying finite-size
effects.

\clearpage

\end{widetext}


\begin{thebibliography}{74}%
\makeatletter
\providecommand \@ifxundefined [1]{%
 \@ifx{#1\undefined}
}%
\providecommand \@ifnum [1]{%
 \ifnum #1\expandafter \@firstoftwo
 \else \expandafter \@secondoftwo
 \fi
}%
\providecommand \@ifx [1]{%
 \ifx #1\expandafter \@firstoftwo
 \else \expandafter \@secondoftwo
 \fi
}%
\providecommand \natexlab [1]{#1}%
\providecommand \enquote  [1]{``#1''}%
\providecommand \bibnamefont  [1]{#1}%
\providecommand \bibfnamefont [1]{#1}%
\providecommand \citenamefont [1]{#1}%
\providecommand \href@noop [0]{\@secondoftwo}%
\providecommand \href [0]{\begingroup \@sanitize@url \@href}%
\providecommand \@href[1]{\@@startlink{#1}\@@href}%
\providecommand \@@href[1]{\endgroup#1\@@endlink}%
\providecommand \@sanitize@url [0]{\catcode `\\12\catcode `\$12\catcode
  `\&12\catcode `\#12\catcode `\^12\catcode `\_12\catcode `\%12\relax}%
\providecommand \@@startlink[1]{}%
\providecommand \@@endlink[0]{}%
\providecommand \url  [0]{\begingroup\@sanitize@url \@url }%
\providecommand \@url [1]{\endgroup\@href {#1}{\urlprefix }}%
\providecommand \urlprefix  [0]{URL }%
\providecommand \Eprint [0]{\href }%
\providecommand \doibase [0]{http://dx.doi.org/}%
\providecommand \selectlanguage [0]{\@gobble}%
\providecommand \bibinfo  [0]{\@secondoftwo}%
\providecommand \bibfield  [0]{\@secondoftwo}%
\providecommand \translation [1]{[#1]}%
\providecommand \BibitemOpen [0]{}%
\providecommand \bibitemStop [0]{}%
\providecommand \bibitemNoStop [0]{.\EOS\space}%
\providecommand \EOS [0]{\spacefactor3000\relax}%
\providecommand \BibitemShut  [1]{\csname bibitem#1\endcsname}%
\let\auto@bib@innerbib\@empty
\bibitem [{\citenamefont {Kitaev}(2006)}]{kitaev2006anyons}%
  \BibitemOpen
  \bibfield  {author} {\bibinfo {author} {\bibfnamefont {A.}~\bibnamefont
  {Kitaev}},\ }\href {\doibase https://doi.org/10.1016/j.aop.2005.10.005}
  {\bibfield  {journal} {\bibinfo  {journal} {Ann. Phys.}\ }\textbf {\bibinfo
  {volume} {321}},\ \bibinfo {pages} {2 } (\bibinfo {year} {2006})}\BibitemShut
  {NoStop}%
\bibitem [{\citenamefont {Balents}(2010)}]{balents2010spin}%
  \BibitemOpen
  \bibfield  {author} {\bibinfo {author} {\bibfnamefont {L.}~\bibnamefont
  {Balents}},\ }\href {https://doi.org/10.1038/nature08917} {\bibfield
  {journal} {\bibinfo  {journal} {Nature}\ }\textbf {\bibinfo {volume} {464}},\
  \bibinfo {pages} {199} (\bibinfo {year} {2010})}\BibitemShut {NoStop}%
\bibitem [{\citenamefont {Savary}\ and\ \citenamefont
  {Balents}(2016)}]{savary2016quantum}%
  \BibitemOpen
  \bibfield  {author} {\bibinfo {author} {\bibfnamefont {L.}~\bibnamefont
  {Savary}}\ and\ \bibinfo {author} {\bibfnamefont {L.}~\bibnamefont
  {Balents}},\ }\href {\doibase 10.1088/0034-4885/80/1/016502} {\bibfield
  {journal} {\bibinfo  {journal} {Rep. Prog. Phys.}\ }\textbf {\bibinfo
  {volume} {80}},\ \bibinfo {pages} {016502} (\bibinfo {year}
  {2016})}\BibitemShut {NoStop}%
\bibitem [{\citenamefont {Jackeli}\ and\ \citenamefont
  {Khaliullin}(2009)}]{jackeli2009mott}%
  \BibitemOpen
  \bibfield  {author} {\bibinfo {author} {\bibfnamefont {G.}~\bibnamefont
  {Jackeli}}\ and\ \bibinfo {author} {\bibfnamefont {G.}~\bibnamefont
  {Khaliullin}},\ }\href {\doibase 10.1103/PhysRevLett.102.017205} {\bibfield
  {journal} {\bibinfo  {journal} {Phys. Rev. Lett.}\ }\textbf {\bibinfo
  {volume} {102}},\ \bibinfo {pages} {017205} (\bibinfo {year}
  {2009})}\BibitemShut {NoStop}%
\bibitem [{\citenamefont {Rau}\ \emph {et~al.}(2016)\citenamefont {Rau},
  \citenamefont {Lee},\ and\ \citenamefont {Kee}}]{rau2016spin}%
  \BibitemOpen
  \bibfield  {author} {\bibinfo {author} {\bibfnamefont {J.~G.}\ \bibnamefont
  {Rau}}, \bibinfo {author} {\bibfnamefont {E.~K.-H.}\ \bibnamefont {Lee}}, \
  and\ \bibinfo {author} {\bibfnamefont {H.-Y.}\ \bibnamefont {Kee}},\ }\href
  {\doibase 10.1146/annurev-conmatphys-031115-011319} {\bibfield  {journal}
  {\bibinfo  {journal} {Annu. Rev. Condens. Matter Phys.}\ }\textbf {\bibinfo
  {volume} {7}},\ \bibinfo {pages} {195} (\bibinfo {year} {2016})}\BibitemShut
  {NoStop}%
\bibitem [{\citenamefont {{Trebst}}(2017)}]{trebst2017kitaev}%
  \BibitemOpen
  \bibfield  {author} {\bibinfo {author} {\bibfnamefont {S.}~\bibnamefont
  {{Trebst}}},\ }\href@noop {} {\bibfield  {journal} {\bibinfo  {journal}
  {ArXiv e-prints}\ } (\bibinfo {year} {2017})},\ \Eprint
  {http://arxiv.org/abs/1701.07056} {arXiv:1701.07056 [cond-mat.str-el]}
  \BibitemShut {NoStop}%
\bibitem [{\citenamefont {Hermanns}\ \emph {et~al.}(2018)\citenamefont
  {Hermanns}, \citenamefont {Kimchi},\ and\ \citenamefont
  {Knolle}}]{hermanns2018physics}%
  \BibitemOpen
  \bibfield  {author} {\bibinfo {author} {\bibfnamefont {M.}~\bibnamefont
  {Hermanns}}, \bibinfo {author} {\bibfnamefont {I.}~\bibnamefont {Kimchi}}, \
  and\ \bibinfo {author} {\bibfnamefont {J.}~\bibnamefont {Knolle}},\ }\href
  {\doibase 10.1146/annurev-conmatphys-033117-053934} {\bibfield  {journal}
  {\bibinfo  {journal} {Annu. Rev. Condens. Matter Phys.}\ }\textbf {\bibinfo
  {volume} {9}},\ \bibinfo {pages} {17} (\bibinfo {year} {2018})}\BibitemShut
  {NoStop}%
\bibitem [{\citenamefont {Singh}\ and\ \citenamefont
  {Gegenwart}(2010)}]{singh2010antiferromagnetic}%
  \BibitemOpen
  \bibfield  {author} {\bibinfo {author} {\bibfnamefont {Y.}~\bibnamefont
  {Singh}}\ and\ \bibinfo {author} {\bibfnamefont {P.}~\bibnamefont
  {Gegenwart}},\ }\href {\doibase 10.1103/PhysRevB.82.064412} {\bibfield
  {journal} {\bibinfo  {journal} {Phys. Rev. B}\ }\textbf {\bibinfo {volume}
  {82}},\ \bibinfo {pages} {064412} (\bibinfo {year} {2010})}\BibitemShut
  {NoStop}%
\bibitem [{\citenamefont {Liu}\ \emph {et~al.}(2011)\citenamefont {Liu},
  \citenamefont {Berlijn}, \citenamefont {Yin}, \citenamefont {Ku},
  \citenamefont {Tsvelik}, \citenamefont {Kim}, \citenamefont {Gretarsson},
  \citenamefont {Singh}, \citenamefont {Gegenwart},\ and\ \citenamefont
  {Hill}}]{liu2011longrange}%
  \BibitemOpen
  \bibfield  {author} {\bibinfo {author} {\bibfnamefont {X.}~\bibnamefont
  {Liu}}, \bibinfo {author} {\bibfnamefont {T.}~\bibnamefont {Berlijn}},
  \bibinfo {author} {\bibfnamefont {W.-G.}\ \bibnamefont {Yin}}, \bibinfo
  {author} {\bibfnamefont {W.}~\bibnamefont {Ku}}, \bibinfo {author}
  {\bibfnamefont {A.}~\bibnamefont {Tsvelik}}, \bibinfo {author} {\bibfnamefont
  {Y.-J.}\ \bibnamefont {Kim}}, \bibinfo {author} {\bibfnamefont
  {H.}~\bibnamefont {Gretarsson}}, \bibinfo {author} {\bibfnamefont
  {Y.}~\bibnamefont {Singh}}, \bibinfo {author} {\bibfnamefont
  {P.}~\bibnamefont {Gegenwart}}, \ and\ \bibinfo {author} {\bibfnamefont
  {J.~P.}\ \bibnamefont {Hill}},\ }\href {\doibase 10.1103/PhysRevB.83.220403}
  {\bibfield  {journal} {\bibinfo  {journal} {Phys. Rev. B}\ }\textbf {\bibinfo
  {volume} {83}},\ \bibinfo {pages} {220403} (\bibinfo {year}
  {2011})}\BibitemShut {NoStop}%
\bibitem [{\citenamefont {Singh}\ \emph {et~al.}(2012)\citenamefont {Singh},
  \citenamefont {Manni}, \citenamefont {Reuther}, \citenamefont {Berlijn},
  \citenamefont {Thomale}, \citenamefont {Ku}, \citenamefont {Trebst},\ and\
  \citenamefont {Gegenwart}}]{singh2012relevance}%
  \BibitemOpen
  \bibfield  {author} {\bibinfo {author} {\bibfnamefont {Y.}~\bibnamefont
  {Singh}}, \bibinfo {author} {\bibfnamefont {S.}~\bibnamefont {Manni}},
  \bibinfo {author} {\bibfnamefont {J.}~\bibnamefont {Reuther}}, \bibinfo
  {author} {\bibfnamefont {T.}~\bibnamefont {Berlijn}}, \bibinfo {author}
  {\bibfnamefont {R.}~\bibnamefont {Thomale}}, \bibinfo {author} {\bibfnamefont
  {W.}~\bibnamefont {Ku}}, \bibinfo {author} {\bibfnamefont {S.}~\bibnamefont
  {Trebst}}, \ and\ \bibinfo {author} {\bibfnamefont {P.}~\bibnamefont
  {Gegenwart}},\ }\href {\doibase 10.1103/PhysRevLett.108.127203} {\bibfield
  {journal} {\bibinfo  {journal} {Phys. Rev. Lett.}\ }\textbf {\bibinfo
  {volume} {108}},\ \bibinfo {pages} {127203} (\bibinfo {year}
  {2012})}\BibitemShut {NoStop}%
\bibitem [{\citenamefont {Choi}\ \emph {et~al.}(2012)\citenamefont {Choi},
  \citenamefont {Coldea}, \citenamefont {Kolmogorov}, \citenamefont
  {Lancaster}, \citenamefont {Mazin}, \citenamefont {Blundell}, \citenamefont
  {Radaelli}, \citenamefont {Singh}, \citenamefont {Gegenwart}, \citenamefont
  {Choi}, \citenamefont {Cheong}, \citenamefont {Baker}, \citenamefont
  {Stock},\ and\ \citenamefont {Taylor}}]{choi2012spin}%
  \BibitemOpen
  \bibfield  {author} {\bibinfo {author} {\bibfnamefont {S.~K.}\ \bibnamefont
  {Choi}}, \bibinfo {author} {\bibfnamefont {R.}~\bibnamefont {Coldea}},
  \bibinfo {author} {\bibfnamefont {A.~N.}\ \bibnamefont {Kolmogorov}},
  \bibinfo {author} {\bibfnamefont {T.}~\bibnamefont {Lancaster}}, \bibinfo
  {author} {\bibfnamefont {I.~I.}\ \bibnamefont {Mazin}}, \bibinfo {author}
  {\bibfnamefont {S.~J.}\ \bibnamefont {Blundell}}, \bibinfo {author}
  {\bibfnamefont {P.~G.}\ \bibnamefont {Radaelli}}, \bibinfo {author}
  {\bibfnamefont {Y.}~\bibnamefont {Singh}}, \bibinfo {author} {\bibfnamefont
  {P.}~\bibnamefont {Gegenwart}}, \bibinfo {author} {\bibfnamefont {K.~R.}\
  \bibnamefont {Choi}}, \bibinfo {author} {\bibfnamefont {S.-W.}\ \bibnamefont
  {Cheong}}, \bibinfo {author} {\bibfnamefont {P.~J.}\ \bibnamefont {Baker}},
  \bibinfo {author} {\bibfnamefont {C.}~\bibnamefont {Stock}}, \ and\ \bibinfo
  {author} {\bibfnamefont {J.}~\bibnamefont {Taylor}},\ }\href {\doibase
  10.1103/PhysRevLett.108.127204} {\bibfield  {journal} {\bibinfo  {journal}
  {Phys. Rev. Lett.}\ }\textbf {\bibinfo {volume} {108}},\ \bibinfo {pages}
  {127204} (\bibinfo {year} {2012})}\BibitemShut {NoStop}%
\bibitem [{\citenamefont {Ye}\ \emph {et~al.}(2012)\citenamefont {Ye},
  \citenamefont {Chi}, \citenamefont {Cao}, \citenamefont {Chakoumakos},
  \citenamefont {Fernandez-Baca}, \citenamefont {Custelcean}, \citenamefont
  {Qi}, \citenamefont {Korneta},\ and\ \citenamefont {Cao}}]{ye2012direct}%
  \BibitemOpen
  \bibfield  {author} {\bibinfo {author} {\bibfnamefont {F.}~\bibnamefont
  {Ye}}, \bibinfo {author} {\bibfnamefont {S.}~\bibnamefont {Chi}}, \bibinfo
  {author} {\bibfnamefont {H.}~\bibnamefont {Cao}}, \bibinfo {author}
  {\bibfnamefont {B.~C.}\ \bibnamefont {Chakoumakos}}, \bibinfo {author}
  {\bibfnamefont {J.~A.}\ \bibnamefont {Fernandez-Baca}}, \bibinfo {author}
  {\bibfnamefont {R.}~\bibnamefont {Custelcean}}, \bibinfo {author}
  {\bibfnamefont {T.~F.}\ \bibnamefont {Qi}}, \bibinfo {author} {\bibfnamefont
  {O.~B.}\ \bibnamefont {Korneta}}, \ and\ \bibinfo {author} {\bibfnamefont
  {G.}~\bibnamefont {Cao}},\ }\href {\doibase 10.1103/PhysRevB.85.180403}
  {\bibfield  {journal} {\bibinfo  {journal} {Phys. Rev. B}\ }\textbf {\bibinfo
  {volume} {85}},\ \bibinfo {pages} {180403} (\bibinfo {year}
  {2012})}\BibitemShut {NoStop}%
\bibitem [{\citenamefont {Comin}\ \emph {et~al.}(2012)\citenamefont {Comin},
  \citenamefont {Levy}, \citenamefont {Ludbrook}, \citenamefont {Zhu},
  \citenamefont {Veenstra}, \citenamefont {Rosen}, \citenamefont {Singh},
  \citenamefont {Gegenwart}, \citenamefont {Stricker}, \citenamefont {Hancock},
  \citenamefont {van~der Marel}, \citenamefont {Elfimov},\ and\ \citenamefont
  {Damascelli}}]{comin2012novel}%
  \BibitemOpen
  \bibfield  {author} {\bibinfo {author} {\bibfnamefont {R.}~\bibnamefont
  {Comin}}, \bibinfo {author} {\bibfnamefont {G.}~\bibnamefont {Levy}},
  \bibinfo {author} {\bibfnamefont {B.}~\bibnamefont {Ludbrook}}, \bibinfo
  {author} {\bibfnamefont {Z.-H.}\ \bibnamefont {Zhu}}, \bibinfo {author}
  {\bibfnamefont {C.~N.}\ \bibnamefont {Veenstra}}, \bibinfo {author}
  {\bibfnamefont {J.~A.}\ \bibnamefont {Rosen}}, \bibinfo {author}
  {\bibfnamefont {Y.}~\bibnamefont {Singh}}, \bibinfo {author} {\bibfnamefont
  {P.}~\bibnamefont {Gegenwart}}, \bibinfo {author} {\bibfnamefont
  {D.}~\bibnamefont {Stricker}}, \bibinfo {author} {\bibfnamefont {J.~N.}\
  \bibnamefont {Hancock}}, \bibinfo {author} {\bibfnamefont {D.}~\bibnamefont
  {van~der Marel}}, \bibinfo {author} {\bibfnamefont {I.~S.}\ \bibnamefont
  {Elfimov}}, \ and\ \bibinfo {author} {\bibfnamefont {A.}~\bibnamefont
  {Damascelli}},\ }\href {\doibase 10.1103/PhysRevLett.109.266406} {\bibfield
  {journal} {\bibinfo  {journal} {Phys. Rev. Lett.}\ }\textbf {\bibinfo
  {volume} {109}},\ \bibinfo {pages} {266406} (\bibinfo {year}
  {2012})}\BibitemShut {NoStop}%
\bibitem [{\citenamefont {Hwan~Chun}\ \emph {et~al.}(2015)\citenamefont
  {Hwan~Chun}, \citenamefont {Kim}, \citenamefont {Kim}, \citenamefont {Zheng},
  \citenamefont {Stoumpos}, \citenamefont {Malliakas}, \citenamefont
  {Mitchell}, \citenamefont {Mehlawat}, \citenamefont {Singh}, \citenamefont
  {Choi}, \citenamefont {Gog}, \citenamefont {Al-Zein}, \citenamefont {Sala},
  \citenamefont {Krisch}, \citenamefont {Chaloupka}, \citenamefont {Jackeli},
  \citenamefont {Khaliullin},\ and\ \citenamefont {Kim}}]{chun2015direct}%
  \BibitemOpen
  \bibfield  {author} {\bibinfo {author} {\bibfnamefont {S.}~\bibnamefont
  {Hwan~Chun}}, \bibinfo {author} {\bibfnamefont {J.-W.}\ \bibnamefont {Kim}},
  \bibinfo {author} {\bibfnamefont {J.}~\bibnamefont {Kim}}, \bibinfo {author}
  {\bibfnamefont {H.}~\bibnamefont {Zheng}}, \bibinfo {author} {\bibfnamefont
  {C.~C.}\ \bibnamefont {Stoumpos}}, \bibinfo {author} {\bibfnamefont {C.~D.}\
  \bibnamefont {Malliakas}}, \bibinfo {author} {\bibfnamefont {J.~F.}\
  \bibnamefont {Mitchell}}, \bibinfo {author} {\bibfnamefont {K.}~\bibnamefont
  {Mehlawat}}, \bibinfo {author} {\bibfnamefont {Y.}~\bibnamefont {Singh}},
  \bibinfo {author} {\bibfnamefont {Y.}~\bibnamefont {Choi}}, \bibinfo {author}
  {\bibfnamefont {T.}~\bibnamefont {Gog}}, \bibinfo {author} {\bibfnamefont
  {A.}~\bibnamefont {Al-Zein}}, \bibinfo {author} {\bibfnamefont {M.~M.}\
  \bibnamefont {Sala}}, \bibinfo {author} {\bibfnamefont {M.}~\bibnamefont
  {Krisch}}, \bibinfo {author} {\bibfnamefont {J.}~\bibnamefont {Chaloupka}},
  \bibinfo {author} {\bibfnamefont {G.}~\bibnamefont {Jackeli}}, \bibinfo
  {author} {\bibfnamefont {G.}~\bibnamefont {Khaliullin}}, \ and\ \bibinfo
  {author} {\bibfnamefont {B.~J.}\ \bibnamefont {Kim}},\ }\href
  {http://dx.doi.org/10.1038/nphys3322} {\bibfield  {journal} {\bibinfo
  {journal} {Nat. Phys.}\ }\textbf {\bibinfo {volume} {11}},\ \bibinfo {pages}
  {462 } (\bibinfo {year} {2015})}\BibitemShut {NoStop}%
\bibitem [{\citenamefont {Williams}\ \emph {et~al.}(2016)\citenamefont
  {Williams}, \citenamefont {Johnson}, \citenamefont {Freund}, \citenamefont
  {Choi}, \citenamefont {Jesche}, \citenamefont {Kimchi}, \citenamefont
  {Manni}, \citenamefont {Bombardi}, \citenamefont {Manuel}, \citenamefont
  {Gegenwart},\ and\ \citenamefont {Coldea}}]{williams2016incommensurate}%
  \BibitemOpen
  \bibfield  {author} {\bibinfo {author} {\bibfnamefont {S.~C.}\ \bibnamefont
  {Williams}}, \bibinfo {author} {\bibfnamefont {R.~D.}\ \bibnamefont
  {Johnson}}, \bibinfo {author} {\bibfnamefont {F.}~\bibnamefont {Freund}},
  \bibinfo {author} {\bibfnamefont {S.}~\bibnamefont {Choi}}, \bibinfo {author}
  {\bibfnamefont {A.}~\bibnamefont {Jesche}}, \bibinfo {author} {\bibfnamefont
  {I.}~\bibnamefont {Kimchi}}, \bibinfo {author} {\bibfnamefont
  {S.}~\bibnamefont {Manni}}, \bibinfo {author} {\bibfnamefont
  {A.}~\bibnamefont {Bombardi}}, \bibinfo {author} {\bibfnamefont
  {P.}~\bibnamefont {Manuel}}, \bibinfo {author} {\bibfnamefont
  {P.}~\bibnamefont {Gegenwart}}, \ and\ \bibinfo {author} {\bibfnamefont
  {R.}~\bibnamefont {Coldea}},\ }\href {\doibase 10.1103/PhysRevB.93.195158}
  {\bibfield  {journal} {\bibinfo  {journal} {Phys. Rev. B}\ }\textbf {\bibinfo
  {volume} {93}},\ \bibinfo {pages} {195158} (\bibinfo {year}
  {2016})}\BibitemShut {NoStop}%
\bibitem [{\citenamefont {Kitagawa}\ \emph {et~al.}(2018)\citenamefont
  {Kitagawa}, \citenamefont {Takayama}, \citenamefont {Matsumoto},
  \citenamefont {Kato}, \citenamefont {Takano}, \citenamefont {Kishimoto},
  \citenamefont {Dinnebier}, \citenamefont {Jackeli},\ and\ \citenamefont
  {Takagi}}]{kitagawa2018spin}%
  \BibitemOpen
  \bibfield  {author} {\bibinfo {author} {\bibfnamefont {K.}~\bibnamefont
  {Kitagawa}}, \bibinfo {author} {\bibfnamefont {T.}~\bibnamefont {Takayama}},
  \bibinfo {author} {\bibfnamefont {Y.}~\bibnamefont {Matsumoto}}, \bibinfo
  {author} {\bibfnamefont {A.}~\bibnamefont {Kato}}, \bibinfo {author}
  {\bibfnamefont {R.}~\bibnamefont {Takano}}, \bibinfo {author} {\bibfnamefont
  {Y.}~\bibnamefont {Kishimoto}}, \bibinfo {author} {\bibfnamefont
  {R.}~\bibnamefont {Dinnebier}}, \bibinfo {author} {\bibfnamefont
  {G.}~\bibnamefont {Jackeli}}, \ and\ \bibinfo {author} {\bibfnamefont
  {H.}~\bibnamefont {Takagi}},\ }\href {\doibase
  http://dx.doi.org/10.1038/nature25482} {\bibfield  {journal} {\bibinfo
  {journal} {Nature}\ }\textbf {\bibinfo {volume} {554}},\ \bibinfo {pages}
  {341} (\bibinfo {year} {2018})}\BibitemShut {NoStop}%
\bibitem [{\citenamefont {Plumb}\ \emph {et~al.}(2014)\citenamefont {Plumb},
  \citenamefont {Clancy}, \citenamefont {Sandilands}, \citenamefont {Shankar},
  \citenamefont {Hu}, \citenamefont {Burch}, \citenamefont {Kee},\ and\
  \citenamefont {Kim}}]{plumb2014spin}%
  \BibitemOpen
  \bibfield  {author} {\bibinfo {author} {\bibfnamefont {K.~W.}\ \bibnamefont
  {Plumb}}, \bibinfo {author} {\bibfnamefont {J.~P.}\ \bibnamefont {Clancy}},
  \bibinfo {author} {\bibfnamefont {L.~J.}\ \bibnamefont {Sandilands}},
  \bibinfo {author} {\bibfnamefont {V.~V.}\ \bibnamefont {Shankar}}, \bibinfo
  {author} {\bibfnamefont {Y.~F.}\ \bibnamefont {Hu}}, \bibinfo {author}
  {\bibfnamefont {K.~S.}\ \bibnamefont {Burch}}, \bibinfo {author}
  {\bibfnamefont {H.-Y.}\ \bibnamefont {Kee}}, \ and\ \bibinfo {author}
  {\bibfnamefont {Y.-J.}\ \bibnamefont {Kim}},\ }\href {\doibase
  10.1103/PhysRevB.90.041112} {\bibfield  {journal} {\bibinfo  {journal} {Phys.
  Rev. B}\ }\textbf {\bibinfo {volume} {90}},\ \bibinfo {pages} {041112}
  (\bibinfo {year} {2014})}\BibitemShut {NoStop}%
\bibitem [{\citenamefont {Sandilands}\ \emph {et~al.}(2015)\citenamefont
  {Sandilands}, \citenamefont {Tian}, \citenamefont {Plumb}, \citenamefont
  {Kim},\ and\ \citenamefont {Burch}}]{sandilands2015scattering}%
  \BibitemOpen
  \bibfield  {author} {\bibinfo {author} {\bibfnamefont {L.~J.}\ \bibnamefont
  {Sandilands}}, \bibinfo {author} {\bibfnamefont {Y.}~\bibnamefont {Tian}},
  \bibinfo {author} {\bibfnamefont {K.~W.}\ \bibnamefont {Plumb}}, \bibinfo
  {author} {\bibfnamefont {Y.-J.}\ \bibnamefont {Kim}}, \ and\ \bibinfo
  {author} {\bibfnamefont {K.~S.}\ \bibnamefont {Burch}},\ }\href {\doibase
  10.1103/PhysRevLett.114.147201} {\bibfield  {journal} {\bibinfo  {journal}
  {Phys. Rev. Lett.}\ }\textbf {\bibinfo {volume} {114}},\ \bibinfo {pages}
  {147201} (\bibinfo {year} {2015})}\BibitemShut {NoStop}%
\bibitem [{\citenamefont {Sears}\ \emph {et~al.}(2015)\citenamefont {Sears},
  \citenamefont {Songvilay}, \citenamefont {Plumb}, \citenamefont {Clancy},
  \citenamefont {Qiu}, \citenamefont {Zhao}, \citenamefont {Parshall},\ and\
  \citenamefont {Kim}}]{sears2015magnetic}%
  \BibitemOpen
  \bibfield  {author} {\bibinfo {author} {\bibfnamefont {J.~A.}\ \bibnamefont
  {Sears}}, \bibinfo {author} {\bibfnamefont {M.}~\bibnamefont {Songvilay}},
  \bibinfo {author} {\bibfnamefont {K.~W.}\ \bibnamefont {Plumb}}, \bibinfo
  {author} {\bibfnamefont {J.~P.}\ \bibnamefont {Clancy}}, \bibinfo {author}
  {\bibfnamefont {Y.}~\bibnamefont {Qiu}}, \bibinfo {author} {\bibfnamefont
  {Y.}~\bibnamefont {Zhao}}, \bibinfo {author} {\bibfnamefont {D.}~\bibnamefont
  {Parshall}}, \ and\ \bibinfo {author} {\bibfnamefont {Y.-J.}\ \bibnamefont
  {Kim}},\ }\href {\doibase 10.1103/PhysRevB.91.144420} {\bibfield  {journal}
  {\bibinfo  {journal} {Phys. Rev. B}\ }\textbf {\bibinfo {volume} {91}},\
  \bibinfo {pages} {144420} (\bibinfo {year} {2015})}\BibitemShut {NoStop}%
\bibitem [{\citenamefont {Majumder}\ \emph {et~al.}(2015)\citenamefont
  {Majumder}, \citenamefont {Schmidt}, \citenamefont {Rosner}, \citenamefont
  {Tsirlin}, \citenamefont {Yasuoka},\ and\ \citenamefont
  {Baenitz}}]{majumder2015anisotropic}%
  \BibitemOpen
  \bibfield  {author} {\bibinfo {author} {\bibfnamefont {M.}~\bibnamefont
  {Majumder}}, \bibinfo {author} {\bibfnamefont {M.}~\bibnamefont {Schmidt}},
  \bibinfo {author} {\bibfnamefont {H.}~\bibnamefont {Rosner}}, \bibinfo
  {author} {\bibfnamefont {A.~A.}\ \bibnamefont {Tsirlin}}, \bibinfo {author}
  {\bibfnamefont {H.}~\bibnamefont {Yasuoka}}, \ and\ \bibinfo {author}
  {\bibfnamefont {M.}~\bibnamefont {Baenitz}},\ }\href {\doibase
  10.1103/PhysRevB.91.180401} {\bibfield  {journal} {\bibinfo  {journal} {Phys.
  Rev. B}\ }\textbf {\bibinfo {volume} {91}},\ \bibinfo {pages} {180401}
  (\bibinfo {year} {2015})}\BibitemShut {NoStop}%
\bibitem [{\citenamefont {Johnson}\ \emph {et~al.}(2015)\citenamefont
  {Johnson}, \citenamefont {Williams}, \citenamefont {Haghighirad},
  \citenamefont {Singleton}, \citenamefont {Zapf}, \citenamefont {Manuel},
  \citenamefont {Mazin}, \citenamefont {Li}, \citenamefont {Jeschke},
  \citenamefont {Valent\'{\i}},\ and\ \citenamefont
  {Coldea}}]{johnson2015monoclinic}%
  \BibitemOpen
  \bibfield  {author} {\bibinfo {author} {\bibfnamefont {R.~D.}\ \bibnamefont
  {Johnson}}, \bibinfo {author} {\bibfnamefont {S.~C.}\ \bibnamefont
  {Williams}}, \bibinfo {author} {\bibfnamefont {A.~A.}\ \bibnamefont
  {Haghighirad}}, \bibinfo {author} {\bibfnamefont {J.}~\bibnamefont
  {Singleton}}, \bibinfo {author} {\bibfnamefont {V.}~\bibnamefont {Zapf}},
  \bibinfo {author} {\bibfnamefont {P.}~\bibnamefont {Manuel}}, \bibinfo
  {author} {\bibfnamefont {I.~I.}\ \bibnamefont {Mazin}}, \bibinfo {author}
  {\bibfnamefont {Y.}~\bibnamefont {Li}}, \bibinfo {author} {\bibfnamefont
  {H.~O.}\ \bibnamefont {Jeschke}}, \bibinfo {author} {\bibfnamefont
  {R.}~\bibnamefont {Valent\'{\i}}}, \ and\ \bibinfo {author} {\bibfnamefont
  {R.}~\bibnamefont {Coldea}},\ }\href {\doibase 10.1103/PhysRevB.92.235119}
  {\bibfield  {journal} {\bibinfo  {journal} {Phys. Rev. B}\ }\textbf {\bibinfo
  {volume} {92}},\ \bibinfo {pages} {235119} (\bibinfo {year}
  {2015})}\BibitemShut {NoStop}%
\bibitem [{\citenamefont {Sandilands}\ \emph {et~al.}(2016)\citenamefont
  {Sandilands}, \citenamefont {Tian}, \citenamefont {Reijnders}, \citenamefont
  {Kim}, \citenamefont {Plumb}, \citenamefont {Kim}, \citenamefont {Kee},\ and\
  \citenamefont {Burch}}]{sandilands2016spin}%
  \BibitemOpen
  \bibfield  {author} {\bibinfo {author} {\bibfnamefont {L.~J.}\ \bibnamefont
  {Sandilands}}, \bibinfo {author} {\bibfnamefont {Y.}~\bibnamefont {Tian}},
  \bibinfo {author} {\bibfnamefont {A.~A.}\ \bibnamefont {Reijnders}}, \bibinfo
  {author} {\bibfnamefont {H.-S.}\ \bibnamefont {Kim}}, \bibinfo {author}
  {\bibfnamefont {K.~W.}\ \bibnamefont {Plumb}}, \bibinfo {author}
  {\bibfnamefont {Y.-J.}\ \bibnamefont {Kim}}, \bibinfo {author} {\bibfnamefont
  {H.-Y.}\ \bibnamefont {Kee}}, \ and\ \bibinfo {author} {\bibfnamefont
  {K.~S.}\ \bibnamefont {Burch}},\ }\href {\doibase 10.1103/PhysRevB.93.075144}
  {\bibfield  {journal} {\bibinfo  {journal} {Phys. Rev. B}\ }\textbf {\bibinfo
  {volume} {93}},\ \bibinfo {pages} {075144} (\bibinfo {year}
  {2016})}\BibitemShut {NoStop}%
\bibitem [{\citenamefont {Banerjee}\ \emph {et~al.}(2016)\citenamefont
  {Banerjee}, \citenamefont {Bridges}, \citenamefont {Yan}, \citenamefont
  {Aczel}, \citenamefont {Li}, \citenamefont {Stone}, \citenamefont {Granroth},
  \citenamefont {Lumsden}, \citenamefont {Yiu}, \citenamefont {Knolle},
  \citenamefont {Bhattacharjee}, \citenamefont {Kovrizhin}, \citenamefont
  {Moessner}, \citenamefont {Tennant}, \citenamefont {G.},\ and\ \citenamefont
  {Nagler}}]{banerjee2016proximate}%
  \BibitemOpen
  \bibfield  {author} {\bibinfo {author} {\bibfnamefont {A.}~\bibnamefont
  {Banerjee}}, \bibinfo {author} {\bibfnamefont {C.~A.}\ \bibnamefont
  {Bridges}}, \bibinfo {author} {\bibfnamefont {J.-Q.}\ \bibnamefont {Yan}},
  \bibinfo {author} {\bibfnamefont {A.~A.}\ \bibnamefont {Aczel}}, \bibinfo
  {author} {\bibfnamefont {L.}~\bibnamefont {Li}}, \bibinfo {author}
  {\bibfnamefont {M.~B.}\ \bibnamefont {Stone}}, \bibinfo {author}
  {\bibfnamefont {G.~E.}\ \bibnamefont {Granroth}}, \bibinfo {author}
  {\bibfnamefont {M.~D.}\ \bibnamefont {Lumsden}}, \bibinfo {author}
  {\bibfnamefont {Y.}~\bibnamefont {Yiu}}, \bibinfo {author} {\bibfnamefont
  {J.}~\bibnamefont {Knolle}}, \bibinfo {author} {\bibfnamefont
  {S.}~\bibnamefont {Bhattacharjee}}, \bibinfo {author} {\bibfnamefont {D.~L.}\
  \bibnamefont {Kovrizhin}}, \bibinfo {author} {\bibfnamefont {R.}~\bibnamefont
  {Moessner}}, \bibinfo {author} {\bibfnamefont {D.~A.}\ \bibnamefont
  {Tennant}}, \bibinfo {author} {\bibfnamefont {M.~D.}\ \bibnamefont {G.}}, \
  and\ \bibinfo {author} {\bibfnamefont {S.~E.}\ \bibnamefont {Nagler}},\
  }\href {\doibase 10.1038/nmat4604} {\bibfield  {journal} {\bibinfo  {journal}
  {Nature materials}\ } (\bibinfo {year} {2016}),\
  10.1038/nmat4604}\BibitemShut {NoStop}%
\bibitem [{\citenamefont {Sears}\ \emph {et~al.}(2017)\citenamefont {Sears},
  \citenamefont {Zhao}, \citenamefont {Xu}, \citenamefont {Lynn},\ and\
  \citenamefont {Kim}}]{sears2017phase}%
  \BibitemOpen
  \bibfield  {author} {\bibinfo {author} {\bibfnamefont {J.~A.}\ \bibnamefont
  {Sears}}, \bibinfo {author} {\bibfnamefont {Y.}~\bibnamefont {Zhao}},
  \bibinfo {author} {\bibfnamefont {Z.}~\bibnamefont {Xu}}, \bibinfo {author}
  {\bibfnamefont {J.~W.}\ \bibnamefont {Lynn}}, \ and\ \bibinfo {author}
  {\bibfnamefont {Y.-J.}\ \bibnamefont {Kim}},\ }\href {\doibase
  10.1103/PhysRevB.95.180411} {\bibfield  {journal} {\bibinfo  {journal} {Phys.
  Rev. B}\ }\textbf {\bibinfo {volume} {95}},\ \bibinfo {pages} {180411}
  (\bibinfo {year} {2017})}\BibitemShut {NoStop}%
\bibitem [{\citenamefont {Banerjee}\ \emph {et~al.}(2017)\citenamefont
  {Banerjee}, \citenamefont {Yan}, \citenamefont {Knolle}, \citenamefont
  {Bridges}, \citenamefont {Stone}, \citenamefont {Lumsden}, \citenamefont
  {Mandrus}, \citenamefont {Tennant}, \citenamefont {Moessner},\ and\
  \citenamefont {Nagler}}]{banerjee2017neutron}%
  \BibitemOpen
  \bibfield  {author} {\bibinfo {author} {\bibfnamefont {A.}~\bibnamefont
  {Banerjee}}, \bibinfo {author} {\bibfnamefont {J.}~\bibnamefont {Yan}},
  \bibinfo {author} {\bibfnamefont {J.}~\bibnamefont {Knolle}}, \bibinfo
  {author} {\bibfnamefont {C.~A.}\ \bibnamefont {Bridges}}, \bibinfo {author}
  {\bibfnamefont {M.~B.}\ \bibnamefont {Stone}}, \bibinfo {author}
  {\bibfnamefont {M.~D.}\ \bibnamefont {Lumsden}}, \bibinfo {author}
  {\bibfnamefont {D.~G.}\ \bibnamefont {Mandrus}}, \bibinfo {author}
  {\bibfnamefont {D.~A.}\ \bibnamefont {Tennant}}, \bibinfo {author}
  {\bibfnamefont {R.}~\bibnamefont {Moessner}}, \ and\ \bibinfo {author}
  {\bibfnamefont {S.~E.}\ \bibnamefont {Nagler}},\ }\href {\doibase
  10.1126/science.aah6015} {\bibfield  {journal} {\bibinfo  {journal}
  {Science}\ }\textbf {\bibinfo {volume} {356}},\ \bibinfo {pages} {1055}
  (\bibinfo {year} {2017})}\BibitemShut {NoStop}%
\bibitem [{\citenamefont {Baek}\ \emph {et~al.}(2017)\citenamefont {Baek},
  \citenamefont {Do}, \citenamefont {Choi}, \citenamefont {Kwon}, \citenamefont
  {Wolter}, \citenamefont {Nishimoto}, \citenamefont {van~den Brink},\ and\
  \citenamefont {B\"uchner}}]{baek2017evidence}%
  \BibitemOpen
  \bibfield  {author} {\bibinfo {author} {\bibfnamefont {S.-H.}\ \bibnamefont
  {Baek}}, \bibinfo {author} {\bibfnamefont {S.-H.}\ \bibnamefont {Do}},
  \bibinfo {author} {\bibfnamefont {K.-Y.}\ \bibnamefont {Choi}}, \bibinfo
  {author} {\bibfnamefont {Y.~S.}\ \bibnamefont {Kwon}}, \bibinfo {author}
  {\bibfnamefont {A.~U.~B.}\ \bibnamefont {Wolter}}, \bibinfo {author}
  {\bibfnamefont {S.}~\bibnamefont {Nishimoto}}, \bibinfo {author}
  {\bibfnamefont {J.}~\bibnamefont {van~den Brink}}, \ and\ \bibinfo {author}
  {\bibfnamefont {B.}~\bibnamefont {B\"uchner}},\ }\href {\doibase
  10.1103/PhysRevLett.119.037201} {\bibfield  {journal} {\bibinfo  {journal}
  {Phys. Rev. Lett.}\ }\textbf {\bibinfo {volume} {119}},\ \bibinfo {pages}
  {037201} (\bibinfo {year} {2017})}\BibitemShut {NoStop}%
\bibitem [{\citenamefont {Do}\ \emph {et~al.}(2017)\citenamefont {Do},
  \citenamefont {Park}, \citenamefont {Yoshitake}, \citenamefont {Nasu},
  \citenamefont {Motome}, \citenamefont {Kwon}, \citenamefont {Adroja},
  \citenamefont {Voneshen}, \citenamefont {Kim}, \citenamefont {Jang},
  \citenamefont {Park}, \citenamefont {Choi},\ and\ \citenamefont
  {Ji}}]{do2017majorana}%
  \BibitemOpen
  \bibfield  {author} {\bibinfo {author} {\bibfnamefont {S.-H.}\ \bibnamefont
  {Do}}, \bibinfo {author} {\bibfnamefont {S.-Y.}\ \bibnamefont {Park}},
  \bibinfo {author} {\bibfnamefont {J.}~\bibnamefont {Yoshitake}}, \bibinfo
  {author} {\bibfnamefont {J.}~\bibnamefont {Nasu}}, \bibinfo {author}
  {\bibfnamefont {Y.}~\bibnamefont {Motome}}, \bibinfo {author} {\bibfnamefont
  {Y.~S.}\ \bibnamefont {Kwon}}, \bibinfo {author} {\bibfnamefont {D.~T.}\
  \bibnamefont {Adroja}}, \bibinfo {author} {\bibfnamefont {D.~J.}\
  \bibnamefont {Voneshen}}, \bibinfo {author} {\bibfnamefont {K.}~\bibnamefont
  {Kim}}, \bibinfo {author} {\bibfnamefont {T.-H.}\ \bibnamefont {Jang}},
  \bibinfo {author} {\bibfnamefont {J.-H.}\ \bibnamefont {Park}}, \bibinfo
  {author} {\bibfnamefont {K.-Y.}\ \bibnamefont {Choi}}, \ and\ \bibinfo
  {author} {\bibfnamefont {S.}~\bibnamefont {Ji}},\ }\href {\doibase
  http://dx.doi.org/10.1038/nphys4264} {\bibfield  {journal} {\bibinfo
  {journal} {Nat. Phys.}\ }\textbf {\bibinfo {volume} {13}},\ \bibinfo {pages}
  {1079} (\bibinfo {year} {2017})}\BibitemShut {NoStop}%
\bibitem [{\citenamefont {{Banerjee}}\ \emph {et~al.}(2018)\citenamefont
  {{Banerjee}}, \citenamefont {{Lampen-Kelley}}, \citenamefont {{Knolle}},
  \citenamefont {{Balz}}, \citenamefont {{Aczel}}, \citenamefont {{Winn}},
  \citenamefont {{Liu}}, \citenamefont {{Pajerowski}}, \citenamefont {{Yan}},
  \citenamefont {{Bridges}}, \citenamefont {{Savici}}, \citenamefont
  {{Chakoumakos}}, \citenamefont {{Lumsden}}, \citenamefont {{Tennant}},
  \citenamefont {{Moessner}}, \citenamefont {{Mandrus}},\ and\ \citenamefont
  {{Nagler}}}]{banerjee2018excitations}%
  \BibitemOpen
  \bibfield  {author} {\bibinfo {author} {\bibfnamefont {A.}~\bibnamefont
  {{Banerjee}}}, \bibinfo {author} {\bibfnamefont {P.}~\bibnamefont
  {{Lampen-Kelley}}}, \bibinfo {author} {\bibfnamefont {J.}~\bibnamefont
  {{Knolle}}}, \bibinfo {author} {\bibfnamefont {C.}~\bibnamefont {{Balz}}},
  \bibinfo {author} {\bibfnamefont {A.~A.}\ \bibnamefont {{Aczel}}}, \bibinfo
  {author} {\bibfnamefont {B.}~\bibnamefont {{Winn}}}, \bibinfo {author}
  {\bibfnamefont {Y.}~\bibnamefont {{Liu}}}, \bibinfo {author} {\bibfnamefont
  {D.}~\bibnamefont {{Pajerowski}}}, \bibinfo {author} {\bibfnamefont
  {J.}~\bibnamefont {{Yan}}}, \bibinfo {author} {\bibfnamefont {C.~A.}\
  \bibnamefont {{Bridges}}}, \bibinfo {author} {\bibfnamefont {A.~T.}\
  \bibnamefont {{Savici}}}, \bibinfo {author} {\bibfnamefont {B.~C.}\
  \bibnamefont {{Chakoumakos}}}, \bibinfo {author} {\bibfnamefont {M.~D.}\
  \bibnamefont {{Lumsden}}}, \bibinfo {author} {\bibfnamefont {D.~A.}\
  \bibnamefont {{Tennant}}}, \bibinfo {author} {\bibfnamefont {R.}~\bibnamefont
  {{Moessner}}}, \bibinfo {author} {\bibfnamefont {D.~G.}\ \bibnamefont
  {{Mandrus}}}, \ and\ \bibinfo {author} {\bibfnamefont {S.~E.}\ \bibnamefont
  {{Nagler}}},\ }\href {\doibase 10.1038/s41535-018-0079-2} {\bibfield
  {journal} {\bibinfo  {journal} {npj Quantum Materials}\ }\textbf {\bibinfo
  {volume} {3}},\ \bibinfo {eid} {8} (\bibinfo {year} {2018})},\ \Eprint
  {http://arxiv.org/abs/1706.07003} {arXiv:1706.07003 [cond-mat.mtrl-sci]}
  \BibitemShut {NoStop}%
\bibitem [{\citenamefont {Hentrich}\ \emph {et~al.}(2018)\citenamefont
  {Hentrich}, \citenamefont {Wolter}, \citenamefont {Zotos}, \citenamefont
  {Brenig}, \citenamefont {Nowak}, \citenamefont {Isaeva}, \citenamefont
  {Doert}, \citenamefont {Banerjee}, \citenamefont {Lampen-Kelley},
  \citenamefont {Mandrus}, \citenamefont {Nagler}, \citenamefont {Sears},
  \citenamefont {Kim}, \citenamefont {B\"uchner},\ and\ \citenamefont
  {Hess}}]{hentrich2018unusual}%
  \BibitemOpen
  \bibfield  {author} {\bibinfo {author} {\bibfnamefont {R.}~\bibnamefont
  {Hentrich}}, \bibinfo {author} {\bibfnamefont {A.~U.~B.}\ \bibnamefont
  {Wolter}}, \bibinfo {author} {\bibfnamefont {X.}~\bibnamefont {Zotos}},
  \bibinfo {author} {\bibfnamefont {W.}~\bibnamefont {Brenig}}, \bibinfo
  {author} {\bibfnamefont {D.}~\bibnamefont {Nowak}}, \bibinfo {author}
  {\bibfnamefont {A.}~\bibnamefont {Isaeva}}, \bibinfo {author} {\bibfnamefont
  {T.}~\bibnamefont {Doert}}, \bibinfo {author} {\bibfnamefont
  {A.}~\bibnamefont {Banerjee}}, \bibinfo {author} {\bibfnamefont
  {P.}~\bibnamefont {Lampen-Kelley}}, \bibinfo {author} {\bibfnamefont {D.~G.}\
  \bibnamefont {Mandrus}}, \bibinfo {author} {\bibfnamefont {S.~E.}\
  \bibnamefont {Nagler}}, \bibinfo {author} {\bibfnamefont {J.}~\bibnamefont
  {Sears}}, \bibinfo {author} {\bibfnamefont {Y.-J.}\ \bibnamefont {Kim}},
  \bibinfo {author} {\bibfnamefont {B.}~\bibnamefont {B\"uchner}}, \ and\
  \bibinfo {author} {\bibfnamefont {C.}~\bibnamefont {Hess}},\ }\href {\doibase
  10.1103/PhysRevLett.120.117204} {\bibfield  {journal} {\bibinfo  {journal}
  {Phys. Rev. Lett.}\ }\textbf {\bibinfo {volume} {120}},\ \bibinfo {pages}
  {117204} (\bibinfo {year} {2018})}\BibitemShut {NoStop}%
\bibitem [{\citenamefont {Kasahara}\ \emph {et~al.}(2018)\citenamefont
  {Kasahara}, \citenamefont {Ohnishi}, \citenamefont {Mizukami}, \citenamefont
  {Tanaka}, \citenamefont {Ma}, \citenamefont {Sugii}, \citenamefont {Kurita},
  \citenamefont {Tanaka}, \citenamefont {Nasu}, \citenamefont {Motome},
  \citenamefont {Shibauchi},\ and\ \citenamefont
  {Matsuda}}]{kasahara2018majorana}%
  \BibitemOpen
  \bibfield  {author} {\bibinfo {author} {\bibfnamefont {Y.}~\bibnamefont
  {Kasahara}}, \bibinfo {author} {\bibfnamefont {T.}~\bibnamefont {Ohnishi}},
  \bibinfo {author} {\bibfnamefont {Y.}~\bibnamefont {Mizukami}}, \bibinfo
  {author} {\bibfnamefont {O.}~\bibnamefont {Tanaka}}, \bibinfo {author}
  {\bibfnamefont {S.}~\bibnamefont {Ma}}, \bibinfo {author} {\bibfnamefont
  {K.}~\bibnamefont {Sugii}}, \bibinfo {author} {\bibfnamefont
  {N.}~\bibnamefont {Kurita}}, \bibinfo {author} {\bibfnamefont
  {H.}~\bibnamefont {Tanaka}}, \bibinfo {author} {\bibfnamefont
  {J.}~\bibnamefont {Nasu}}, \bibinfo {author} {\bibfnamefont {Y.}~\bibnamefont
  {Motome}}, \bibinfo {author} {\bibfnamefont {T.}~\bibnamefont {Shibauchi}}, \
  and\ \bibinfo {author} {\bibfnamefont {Y.}~\bibnamefont {Matsuda}},\ }\href
  {\doibase 10.1038/s41586-018-0274-0} {\bibfield  {journal} {\bibinfo
  {journal} {Nature}\ }\textbf {\bibinfo {volume} {559}},\ \bibinfo {pages}
  {227} (\bibinfo {year} {2018})}\BibitemShut {NoStop}%
\bibitem [{\citenamefont {Chaloupka}\ \emph {et~al.}(2010)\citenamefont
  {Chaloupka}, \citenamefont {Jackeli},\ and\ \citenamefont
  {Khaliullin}}]{chaloupka2010kitaev}%
  \BibitemOpen
  \bibfield  {author} {\bibinfo {author} {\bibfnamefont {J.}~\bibnamefont
  {Chaloupka}}, \bibinfo {author} {\bibfnamefont {G.}~\bibnamefont {Jackeli}},
  \ and\ \bibinfo {author} {\bibfnamefont {G.}~\bibnamefont {Khaliullin}},\
  }\href {\doibase 10.1103/PhysRevLett.105.027204} {\bibfield  {journal}
  {\bibinfo  {journal} {Phys. Rev. Lett.}\ }\textbf {\bibinfo {volume} {105}},\
  \bibinfo {pages} {027204} (\bibinfo {year} {2010})}\BibitemShut {NoStop}%
\bibitem [{\citenamefont {Jiang}\ \emph {et~al.}(2011)\citenamefont {Jiang},
  \citenamefont {Gu}, \citenamefont {Qi},\ and\ \citenamefont
  {Trebst}}]{jiang2011possible}%
  \BibitemOpen
  \bibfield  {author} {\bibinfo {author} {\bibfnamefont {H.-C.}\ \bibnamefont
  {Jiang}}, \bibinfo {author} {\bibfnamefont {Z.-C.}\ \bibnamefont {Gu}},
  \bibinfo {author} {\bibfnamefont {X.-L.}\ \bibnamefont {Qi}}, \ and\ \bibinfo
  {author} {\bibfnamefont {S.}~\bibnamefont {Trebst}},\ }\href {\doibase
  10.1103/PhysRevB.83.245104} {\bibfield  {journal} {\bibinfo  {journal} {Phys.
  Rev. B}\ }\textbf {\bibinfo {volume} {83}},\ \bibinfo {pages} {245104}
  (\bibinfo {year} {2011})}\BibitemShut {NoStop}%
\bibitem [{\citenamefont {Reuther}\ \emph {et~al.}(2011)\citenamefont
  {Reuther}, \citenamefont {Thomale},\ and\ \citenamefont
  {Trebst}}]{reuther2011finite}%
  \BibitemOpen
  \bibfield  {author} {\bibinfo {author} {\bibfnamefont {J.}~\bibnamefont
  {Reuther}}, \bibinfo {author} {\bibfnamefont {R.}~\bibnamefont {Thomale}}, \
  and\ \bibinfo {author} {\bibfnamefont {S.}~\bibnamefont {Trebst}},\ }\href
  {\doibase 10.1103/PhysRevB.84.100406} {\bibfield  {journal} {\bibinfo
  {journal} {Phys. Rev. B}\ }\textbf {\bibinfo {volume} {84}},\ \bibinfo
  {pages} {100406} (\bibinfo {year} {2011})}\BibitemShut {NoStop}%
\bibitem [{\citenamefont {Price}\ and\ \citenamefont
  {Perkins}(2012)}]{price2012critical}%
  \BibitemOpen
  \bibfield  {author} {\bibinfo {author} {\bibfnamefont {C.~C.}\ \bibnamefont
  {Price}}\ and\ \bibinfo {author} {\bibfnamefont {N.~B.}\ \bibnamefont
  {Perkins}},\ }\href {\doibase 10.1103/PhysRevLett.109.187201} {\bibfield
  {journal} {\bibinfo  {journal} {Phys. Rev. Lett.}\ }\textbf {\bibinfo
  {volume} {109}},\ \bibinfo {pages} {187201} (\bibinfo {year}
  {2012})}\BibitemShut {NoStop}%
\bibitem [{\citenamefont {Rau}\ \emph {et~al.}(2014)\citenamefont {Rau},
  \citenamefont {Lee},\ and\ \citenamefont {Kee}}]{rau2014generic}%
  \BibitemOpen
  \bibfield  {author} {\bibinfo {author} {\bibfnamefont {J.~G.}\ \bibnamefont
  {Rau}}, \bibinfo {author} {\bibfnamefont {E.~K.-H.}\ \bibnamefont {Lee}}, \
  and\ \bibinfo {author} {\bibfnamefont {H.-Y.}\ \bibnamefont {Kee}},\ }\href
  {\doibase 10.1103/PhysRevLett.112.077204} {\bibfield  {journal} {\bibinfo
  {journal} {Phys. Rev. Lett.}\ }\textbf {\bibinfo {volume} {112}},\ \bibinfo
  {pages} {077204} (\bibinfo {year} {2014})}\BibitemShut {NoStop}%
\bibitem [{\citenamefont {Yamaji}\ \emph {et~al.}(2014)\citenamefont {Yamaji},
  \citenamefont {Nomura}, \citenamefont {Kurita}, \citenamefont {Arita},\ and\
  \citenamefont {Imada}}]{yamaji2014first}%
  \BibitemOpen
  \bibfield  {author} {\bibinfo {author} {\bibfnamefont {Y.}~\bibnamefont
  {Yamaji}}, \bibinfo {author} {\bibfnamefont {Y.}~\bibnamefont {Nomura}},
  \bibinfo {author} {\bibfnamefont {M.}~\bibnamefont {Kurita}}, \bibinfo
  {author} {\bibfnamefont {R.}~\bibnamefont {Arita}}, \ and\ \bibinfo {author}
  {\bibfnamefont {M.}~\bibnamefont {Imada}},\ }\href {\doibase
  10.1103/PhysRevLett.113.107201} {\bibfield  {journal} {\bibinfo  {journal}
  {Phys. Rev. Lett.}\ }\textbf {\bibinfo {volume} {113}},\ \bibinfo {pages}
  {107201} (\bibinfo {year} {2014})}\BibitemShut {NoStop}%
\bibitem [{\citenamefont {Sizyuk}\ \emph {et~al.}(2014)\citenamefont {Sizyuk},
  \citenamefont {Price}, \citenamefont {W\"olfle},\ and\ \citenamefont
  {Perkins}}]{sizyuk2014importance}%
  \BibitemOpen
  \bibfield  {author} {\bibinfo {author} {\bibfnamefont {Y.}~\bibnamefont
  {Sizyuk}}, \bibinfo {author} {\bibfnamefont {C.}~\bibnamefont {Price}},
  \bibinfo {author} {\bibfnamefont {P.}~\bibnamefont {W\"olfle}}, \ and\
  \bibinfo {author} {\bibfnamefont {N.~B.}\ \bibnamefont {Perkins}},\ }\href
  {\doibase 10.1103/PhysRevB.90.155126} {\bibfield  {journal} {\bibinfo
  {journal} {Phys. Rev. B}\ }\textbf {\bibinfo {volume} {90}},\ \bibinfo
  {pages} {155126} (\bibinfo {year} {2014})}\BibitemShut {NoStop}%
\bibitem [{\citenamefont {Sela}\ \emph {et~al.}(2014)\citenamefont {Sela},
  \citenamefont {Jiang}, \citenamefont {Gerlach},\ and\ \citenamefont
  {Trebst}}]{sela2014}%
  \BibitemOpen
  \bibfield  {author} {\bibinfo {author} {\bibfnamefont {E.}~\bibnamefont
  {Sela}}, \bibinfo {author} {\bibfnamefont {H.-C.}\ \bibnamefont {Jiang}},
  \bibinfo {author} {\bibfnamefont {M.~H.}\ \bibnamefont {Gerlach}}, \ and\
  \bibinfo {author} {\bibfnamefont {S.}~\bibnamefont {Trebst}},\ }\href@noop {}
  {\bibfield  {journal} {\bibinfo  {journal} {Physical Review B}\ }\textbf
  {\bibinfo {volume} {90}},\ \bibinfo {pages} {035113} (\bibinfo {year}
  {2014})}\BibitemShut {NoStop}%
\bibitem [{\citenamefont {Rousochatzakis}\ \emph {et~al.}(2015)\citenamefont
  {Rousochatzakis}, \citenamefont {Reuther}, \citenamefont {Thomale},
  \citenamefont {Rachel},\ and\ \citenamefont
  {Perkins}}]{rousochatzakis2015phase}%
  \BibitemOpen
  \bibfield  {author} {\bibinfo {author} {\bibfnamefont {I.}~\bibnamefont
  {Rousochatzakis}}, \bibinfo {author} {\bibfnamefont {J.}~\bibnamefont
  {Reuther}}, \bibinfo {author} {\bibfnamefont {R.}~\bibnamefont {Thomale}},
  \bibinfo {author} {\bibfnamefont {S.}~\bibnamefont {Rachel}}, \ and\ \bibinfo
  {author} {\bibfnamefont {N.~B.}\ \bibnamefont {Perkins}},\ }\href {\doibase
  10.1103/PhysRevX.5.041035} {\bibfield  {journal} {\bibinfo  {journal} {Phys.
  Rev. X}\ }\textbf {\bibinfo {volume} {5}},\ \bibinfo {pages} {041035}
  (\bibinfo {year} {2015})}\BibitemShut {NoStop}%
\bibitem [{\citenamefont {Kim}\ \emph {et~al.}(2015)\citenamefont {Kim},
  \citenamefont {V.}, \citenamefont {Catuneanu},\ and\ \citenamefont
  {Kee}}]{kim2015}%
  \BibitemOpen
  \bibfield  {author} {\bibinfo {author} {\bibfnamefont {H.-S.}\ \bibnamefont
  {Kim}}, \bibinfo {author} {\bibfnamefont {V.}~\bibnamefont {V.}}, \bibinfo
  {author} {\bibfnamefont {A.}~\bibnamefont {Catuneanu}}, \ and\ \bibinfo
  {author} {\bibfnamefont {H.-Y.}\ \bibnamefont {Kee}},\ }\href {\doibase
  10.1103/PhysRevB.91.241110} {\bibfield  {journal} {\bibinfo  {journal} {Phys.
  Rev. B}\ }\textbf {\bibinfo {volume} {91}},\ \bibinfo {pages} {241110}
  (\bibinfo {year} {2015})}\BibitemShut {NoStop}%
\bibitem [{\citenamefont {Yadav}\ \emph {et~al.}(2016)\citenamefont {Yadav},
  \citenamefont {Bogdanov}, \citenamefont {Katukuri}, \citenamefont
  {Nishimoto}, \citenamefont {Van Den~Brink},\ and\ \citenamefont
  {Hozoi}}]{yadav2016}%
  \BibitemOpen
  \bibfield  {author} {\bibinfo {author} {\bibfnamefont {R.}~\bibnamefont
  {Yadav}}, \bibinfo {author} {\bibfnamefont {N.~A.}\ \bibnamefont {Bogdanov}},
  \bibinfo {author} {\bibfnamefont {V.~M.}\ \bibnamefont {Katukuri}}, \bibinfo
  {author} {\bibfnamefont {S.}~\bibnamefont {Nishimoto}}, \bibinfo {author}
  {\bibfnamefont {J.}~\bibnamefont {Van Den~Brink}}, \ and\ \bibinfo {author}
  {\bibfnamefont {L.}~\bibnamefont {Hozoi}},\ }\href@noop {} {\bibfield
  {journal} {\bibinfo  {journal} {Scientific reports}\ }\textbf {\bibinfo
  {volume} {6}},\ \bibinfo {pages} {37925} (\bibinfo {year}
  {2016})}\BibitemShut {NoStop}%
\bibitem [{\citenamefont {Kim}\ and\ \citenamefont {Kee}(2016)}]{kim2016}%
  \BibitemOpen
  \bibfield  {author} {\bibinfo {author} {\bibfnamefont {H.-S.}\ \bibnamefont
  {Kim}}\ and\ \bibinfo {author} {\bibfnamefont {H.-Y.}\ \bibnamefont {Kee}},\
  }\href@noop {} {\bibfield  {journal} {\bibinfo  {journal} {Physical Review
  B}\ }\textbf {\bibinfo {volume} {93}},\ \bibinfo {pages} {155143} (\bibinfo
  {year} {2016})}\BibitemShut {NoStop}%
\bibitem [{\citenamefont {Winter}\ \emph {et~al.}(2016)\citenamefont {Winter},
  \citenamefont {Li}, \citenamefont {Jeschke},\ and\ \citenamefont
  {Valent\'{\i}}}]{winter2016challenges}%
  \BibitemOpen
  \bibfield  {author} {\bibinfo {author} {\bibfnamefont {S.~M.}\ \bibnamefont
  {Winter}}, \bibinfo {author} {\bibfnamefont {Y.}~\bibnamefont {Li}}, \bibinfo
  {author} {\bibfnamefont {H.~O.}\ \bibnamefont {Jeschke}}, \ and\ \bibinfo
  {author} {\bibfnamefont {R.}~\bibnamefont {Valent\'{\i}}},\ }\href {\doibase
  10.1103/PhysRevB.93.214431} {\bibfield  {journal} {\bibinfo  {journal} {Phys.
  Rev. B}\ }\textbf {\bibinfo {volume} {93}},\ \bibinfo {pages} {214431}
  (\bibinfo {year} {2016})}\BibitemShut {NoStop}%
\bibitem [{\citenamefont {Kim}\ \emph {et~al.}(2016)\citenamefont {Kim},
  \citenamefont {Shirakawa},\ and\ \citenamefont
  {Yunoki}}]{kim2016quasimolecular}%
  \BibitemOpen
  \bibfield  {author} {\bibinfo {author} {\bibfnamefont {B.~H.}\ \bibnamefont
  {Kim}}, \bibinfo {author} {\bibfnamefont {T.}~\bibnamefont {Shirakawa}}, \
  and\ \bibinfo {author} {\bibfnamefont {S.}~\bibnamefont {Yunoki}},\
  }\href@noop {} {\bibfield  {journal} {\bibinfo  {journal} {Physical review
  letters}\ }\textbf {\bibinfo {volume} {117}},\ \bibinfo {pages} {187201}
  (\bibinfo {year} {2016})}\BibitemShut {NoStop}%
\bibitem [{\citenamefont {Janssen}\ \emph {et~al.}(2016)\citenamefont
  {Janssen}, \citenamefont {Andrade},\ and\ \citenamefont
  {Vojta}}]{janssen2016honeycomb}%
  \BibitemOpen
  \bibfield  {author} {\bibinfo {author} {\bibfnamefont {L.}~\bibnamefont
  {Janssen}}, \bibinfo {author} {\bibfnamefont {E.~C.}\ \bibnamefont
  {Andrade}}, \ and\ \bibinfo {author} {\bibfnamefont {M.}~\bibnamefont
  {Vojta}},\ }\href@noop {} {\bibfield  {journal} {\bibinfo  {journal}
  {Physical Review Letters}\ }\textbf {\bibinfo {volume} {117}},\ \bibinfo
  {pages} {277202} (\bibinfo {year} {2016})}\BibitemShut {NoStop}%
\bibitem [{\citenamefont {Hou}\ \emph {et~al.}(2017)\citenamefont {Hou},
  \citenamefont {Xiang},\ and\ \citenamefont {Gong}}]{Hou2017}%
  \BibitemOpen
  \bibfield  {author} {\bibinfo {author} {\bibfnamefont {Y.~S.}\ \bibnamefont
  {Hou}}, \bibinfo {author} {\bibfnamefont {H.~J.}\ \bibnamefont {Xiang}}, \
  and\ \bibinfo {author} {\bibfnamefont {X.~G.}\ \bibnamefont {Gong}},\ }\href
  {\doibase 10.1103/PhysRevB.96.054410} {\bibfield  {journal} {\bibinfo
  {journal} {Phys. Rev. B}\ }\textbf {\bibinfo {volume} {96}},\ \bibinfo
  {pages} {054410} (\bibinfo {year} {2017})}\BibitemShut {NoStop}%
\bibitem [{\citenamefont {Winter}\ \emph {et~al.}(2017)\citenamefont {Winter},
  \citenamefont {Tsirlin}, \citenamefont {Daghofer}, \citenamefont {van~den
  Brink}, \citenamefont {Singh}, \citenamefont {Gegenwart},\ and\ \citenamefont
  {Valent{\'\i}}}]{winter2017models}%
  \BibitemOpen
  \bibfield  {author} {\bibinfo {author} {\bibfnamefont {S.~M.}\ \bibnamefont
  {Winter}}, \bibinfo {author} {\bibfnamefont {A.~A.}\ \bibnamefont {Tsirlin}},
  \bibinfo {author} {\bibfnamefont {M.}~\bibnamefont {Daghofer}}, \bibinfo
  {author} {\bibfnamefont {J.}~\bibnamefont {van~den Brink}}, \bibinfo {author}
  {\bibfnamefont {Y.}~\bibnamefont {Singh}}, \bibinfo {author} {\bibfnamefont
  {P.}~\bibnamefont {Gegenwart}}, \ and\ \bibinfo {author} {\bibfnamefont
  {R.}~\bibnamefont {Valent{\'\i}}},\ }\href
  {http://stacks.iop.org/0953-8984/29/i=49/a=493002} {\bibfield  {journal}
  {\bibinfo  {journal} {Journal of Physics: Condensed Matter}\ }\textbf
  {\bibinfo {volume} {29}},\ \bibinfo {pages} {493002} (\bibinfo {year}
  {2017})}\BibitemShut {NoStop}%
\bibitem [{\citenamefont {Samarakoon}\ \emph {et~al.}(2017)\citenamefont
  {Samarakoon}, \citenamefont {Banerjee}, \citenamefont {Zhang}, \citenamefont
  {Kamiya}, \citenamefont {Nagler}, \citenamefont {Tennant}, \citenamefont
  {Lee},\ and\ \citenamefont {Batista}}]{Samarakoon2017}%
  \BibitemOpen
  \bibfield  {author} {\bibinfo {author} {\bibfnamefont {A.~M.}\ \bibnamefont
  {Samarakoon}}, \bibinfo {author} {\bibfnamefont {A.}~\bibnamefont
  {Banerjee}}, \bibinfo {author} {\bibfnamefont {S.-S.}\ \bibnamefont {Zhang}},
  \bibinfo {author} {\bibfnamefont {Y.}~\bibnamefont {Kamiya}}, \bibinfo
  {author} {\bibfnamefont {S.~E.}\ \bibnamefont {Nagler}}, \bibinfo {author}
  {\bibfnamefont {D.~A.}\ \bibnamefont {Tennant}}, \bibinfo {author}
  {\bibfnamefont {S.-H.}\ \bibnamefont {Lee}}, \ and\ \bibinfo {author}
  {\bibfnamefont {C.~D.}\ \bibnamefont {Batista}},\ }\href {\doibase
  10.1103/PhysRevB.96.134408} {\bibfield  {journal} {\bibinfo  {journal} {Phys.
  Rev. B}\ }\textbf {\bibinfo {volume} {96}},\ \bibinfo {pages} {134408}
  (\bibinfo {year} {2017})}\BibitemShut {NoStop}%
\bibitem [{\citenamefont {Ran}\ \emph {et~al.}(2017)\citenamefont {Ran},
  \citenamefont {Wang}, \citenamefont {Wang}, \citenamefont {Dong},
  \citenamefont {Ren}, \citenamefont {Bao}, \citenamefont {Li}, \citenamefont
  {Ma}, \citenamefont {Gan}, \citenamefont {Zhang} \emph {et~al.}}]{Ran2017}%
  \BibitemOpen
  \bibfield  {author} {\bibinfo {author} {\bibfnamefont {K.}~\bibnamefont
  {Ran}}, \bibinfo {author} {\bibfnamefont {J.}~\bibnamefont {Wang}}, \bibinfo
  {author} {\bibfnamefont {W.}~\bibnamefont {Wang}}, \bibinfo {author}
  {\bibfnamefont {Z.-Y.}\ \bibnamefont {Dong}}, \bibinfo {author}
  {\bibfnamefont {X.}~\bibnamefont {Ren}}, \bibinfo {author} {\bibfnamefont
  {S.}~\bibnamefont {Bao}}, \bibinfo {author} {\bibfnamefont {S.}~\bibnamefont
  {Li}}, \bibinfo {author} {\bibfnamefont {Z.}~\bibnamefont {Ma}}, \bibinfo
  {author} {\bibfnamefont {Y.}~\bibnamefont {Gan}}, \bibinfo {author}
  {\bibfnamefont {Y.}~\bibnamefont {Zhang}},  \emph {et~al.},\ }\href@noop {}
  {\bibfield  {journal} {\bibinfo  {journal} {Physical Review Letters}\
  }\textbf {\bibinfo {volume} {118}},\ \bibinfo {pages} {107203} (\bibinfo
  {year} {2017})}\BibitemShut {NoStop}%
\bibitem [{\citenamefont {Samarakoon}\ \emph {et~al.}(2018)\citenamefont
  {Samarakoon}, \citenamefont {Wachtel}, \citenamefont {Yamaji}, \citenamefont
  {Tennant}, \citenamefont {Batista},\ and\ \citenamefont
  {Kim}}]{Samarakoon2018}%
  \BibitemOpen
  \bibfield  {author} {\bibinfo {author} {\bibfnamefont {A.~M.}\ \bibnamefont
  {Samarakoon}}, \bibinfo {author} {\bibfnamefont {G.}~\bibnamefont {Wachtel}},
  \bibinfo {author} {\bibfnamefont {Y.}~\bibnamefont {Yamaji}}, \bibinfo
  {author} {\bibfnamefont {D.~A.}\ \bibnamefont {Tennant}}, \bibinfo {author}
  {\bibfnamefont {C.~D.}\ \bibnamefont {Batista}}, \ and\ \bibinfo {author}
  {\bibfnamefont {Y.~B.}\ \bibnamefont {Kim}},\ }\href {\doibase
  10.1103/PhysRevB.98.045121} {\bibfield  {journal} {\bibinfo  {journal} {Phys.
  Rev. B}\ }\textbf {\bibinfo {volume} {98}},\ \bibinfo {pages} {045121}
  (\bibinfo {year} {2018})}\BibitemShut {NoStop}%
\bibitem [{\citenamefont {{Gordon}}\ \emph {et~al.}(2019)\citenamefont
  {{Gordon}}, \citenamefont {{Catuneanu}}, \citenamefont {{S{\o}rensen}},\ and\
  \citenamefont {{Kee}}}]{gordon2019theory}%
  \BibitemOpen
  \bibfield  {author} {\bibinfo {author} {\bibfnamefont {J.~S.}\ \bibnamefont
  {{Gordon}}}, \bibinfo {author} {\bibfnamefont {A.}~\bibnamefont
  {{Catuneanu}}}, \bibinfo {author} {\bibfnamefont {E.~S.}\ \bibnamefont
  {{S{\o}rensen}}}, \ and\ \bibinfo {author} {\bibfnamefont {H.-Y.}\
  \bibnamefont {{Kee}}},\ }\href@noop {} {\bibfield  {journal} {\bibinfo
  {journal} {arXiv e-prints}\ ,\ \bibinfo {eid} {arXiv:1901.09943}} (\bibinfo
  {year} {2019})},\ \Eprint {http://arxiv.org/abs/1901.09943} {arXiv:1901.09943
  [cond-mat.str-el]} \BibitemShut {NoStop}%
\bibitem [{\citenamefont {Lu}\ and\ \citenamefont {Ran}(2011)}]{lu2011z2}%
  \BibitemOpen
  \bibfield  {author} {\bibinfo {author} {\bibfnamefont {Y.-M.}\ \bibnamefont
  {Lu}}\ and\ \bibinfo {author} {\bibfnamefont {Y.}~\bibnamefont {Ran}},\
  }\href {\doibase 10.1103/PhysRevB.84.024420} {\bibfield  {journal} {\bibinfo
  {journal} {Phys. Rev. B}\ }\textbf {\bibinfo {volume} {84}},\ \bibinfo
  {pages} {024420} (\bibinfo {year} {2011})}\BibitemShut {NoStop}%
\bibitem [{\citenamefont {You}\ \emph {et~al.}(2012)\citenamefont {You},
  \citenamefont {Kimchi},\ and\ \citenamefont {Vishwanath}}]{you2012doping}%
  \BibitemOpen
  \bibfield  {author} {\bibinfo {author} {\bibfnamefont {Y.-Z.}\ \bibnamefont
  {You}}, \bibinfo {author} {\bibfnamefont {I.}~\bibnamefont {Kimchi}}, \ and\
  \bibinfo {author} {\bibfnamefont {A.}~\bibnamefont {Vishwanath}},\ }\href
  {\doibase 10.1103/PhysRevB.86.085145} {\bibfield  {journal} {\bibinfo
  {journal} {Phys. Rev. B}\ }\textbf {\bibinfo {volume} {86}},\ \bibinfo
  {pages} {085145} (\bibinfo {year} {2012})}\BibitemShut {NoStop}%
\bibitem [{\citenamefont {Savary}\ and\ \citenamefont
  {Balents}(2012)}]{savary2012coulombic}%
  \BibitemOpen
  \bibfield  {author} {\bibinfo {author} {\bibfnamefont {L.}~\bibnamefont
  {Savary}}\ and\ \bibinfo {author} {\bibfnamefont {L.}~\bibnamefont
  {Balents}},\ }\href {\doibase 10.1103/PhysRevLett.108.037202} {\bibfield
  {journal} {\bibinfo  {journal} {Phys. Rev. Lett.}\ }\textbf {\bibinfo
  {volume} {108}},\ \bibinfo {pages} {037202} (\bibinfo {year}
  {2012})}\BibitemShut {NoStop}%
\bibitem [{\citenamefont {Lieb}(1994)}]{Lieb94}%
  \BibitemOpen
  \bibfield  {author} {\bibinfo {author} {\bibfnamefont {E.~H.}\ \bibnamefont
  {Lieb}},\ }\href {\doibase 10.1103/PhysRevLett.73.2158} {\bibfield  {journal}
  {\bibinfo  {journal} {Phys. Rev. Lett.}\ }\textbf {\bibinfo {volume} {73}},\
  \bibinfo {pages} {2158} (\bibinfo {year} {1994})}\BibitemShut {NoStop}%
\bibitem [{\citenamefont {Ozawa}\ \emph {et~al.}(2016)\citenamefont {Ozawa},
  \citenamefont {Hayami}, \citenamefont {Barros}, \citenamefont {Chern},
  \citenamefont {Motome},\ and\ \citenamefont {Batista}}]{Ozawa16}%
  \BibitemOpen
  \bibfield  {author} {\bibinfo {author} {\bibfnamefont {R.}~\bibnamefont
  {Ozawa}}, \bibinfo {author} {\bibfnamefont {S.}~\bibnamefont {Hayami}},
  \bibinfo {author} {\bibfnamefont {K.}~\bibnamefont {Barros}}, \bibinfo
  {author} {\bibfnamefont {G.-W.}\ \bibnamefont {Chern}}, \bibinfo {author}
  {\bibfnamefont {Y.}~\bibnamefont {Motome}}, \ and\ \bibinfo {author}
  {\bibfnamefont {C.~D.}\ \bibnamefont {Batista}},\ }\href {\doibase
  10.7566/JPSJ.85.103703} {\bibfield  {journal} {\bibinfo  {journal} {J. Phys.
  Soc. Jpn.}\ }\textbf {\bibinfo {volume} {85}},\ \bibinfo {pages} {103703}
  (\bibinfo {year} {2016})}\BibitemShut {NoStop}%
\bibitem [{\citenamefont {Batista}\ \emph {et~al.}(2016)\citenamefont
  {Batista}, \citenamefont {Lin}, \citenamefont {Hayami},\ and\ \citenamefont
  {Kamiya}}]{Batista16}%
  \BibitemOpen
  \bibfield  {author} {\bibinfo {author} {\bibfnamefont {C.~D.}\ \bibnamefont
  {Batista}}, \bibinfo {author} {\bibfnamefont {S.-Z.}\ \bibnamefont {Lin}},
  \bibinfo {author} {\bibfnamefont {S.}~\bibnamefont {Hayami}}, \ and\ \bibinfo
  {author} {\bibfnamefont {Y.}~\bibnamefont {Kamiya}},\ }\href
  {http://stacks.iop.org/0034-4885/79/i=8/a=084504} {\bibfield  {journal}
  {\bibinfo  {journal} {Reports on Progress in Physics}\ }\textbf {\bibinfo
  {volume} {79}},\ \bibinfo {pages} {084504} (\bibinfo {year}
  {2016})}\BibitemShut {NoStop}%
\bibitem [{\citenamefont {Ozawa}\ \emph {et~al.}(2017)\citenamefont {Ozawa},
  \citenamefont {Hayami},\ and\ \citenamefont {Motome}}]{Ozawa17}%
  \BibitemOpen
  \bibfield  {author} {\bibinfo {author} {\bibfnamefont {R.}~\bibnamefont
  {Ozawa}}, \bibinfo {author} {\bibfnamefont {S.}~\bibnamefont {Hayami}}, \
  and\ \bibinfo {author} {\bibfnamefont {Y.}~\bibnamefont {Motome}},\ }\href
  {\doibase 10.1103/PhysRevLett.118.147205} {\bibfield  {journal} {\bibinfo
  {journal} {Phys. Rev. Lett.}\ }\textbf {\bibinfo {volume} {118}},\ \bibinfo
  {pages} {147205} (\bibinfo {year} {2017})}\BibitemShut {NoStop}%
\bibitem [{not()}]{note0}%
  \BibitemOpen
  \href@noop {} {}\bibinfo {note} {In this Letter, ``$n$-th-neighbor'' means
  that the shortest path connecting the two sites consists of $n$
  bonds.}\BibitemShut {Stop}%
\bibitem [{sup()}]{supp}%
  \BibitemOpen
  \href@noop {} {}\bibinfo {note} {See Supplemental Material for extended
  descriptions of the quadratic Majorana problems and the corresponding nodal
  structures in the various phases, for detailed results on the magnetic
  Friedel oscillations and the dynamical spin structure factor, as well as for
  implementation details of the Monte Carlo simulations.}\BibitemShut {Stop}%
\bibitem [{\citenamefont {Song}\ \emph {et~al.}(2016)\citenamefont {Song},
  \citenamefont {You},\ and\ \citenamefont {Balents}}]{song2016low}%
  \BibitemOpen
  \bibfield  {author} {\bibinfo {author} {\bibfnamefont {X.-Y.}\ \bibnamefont
  {Song}}, \bibinfo {author} {\bibfnamefont {Y.-Z.}\ \bibnamefont {You}}, \
  and\ \bibinfo {author} {\bibfnamefont {L.}~\bibnamefont {Balents}},\ }\href
  {\doibase 10.1103/PhysRevLett.117.037209} {\bibfield  {journal} {\bibinfo
  {journal} {Phys. Rev. Lett.}\ }\textbf {\bibinfo {volume} {117}},\ \bibinfo
  {pages} {037209} (\bibinfo {year} {2016})}\BibitemShut {NoStop}%
\bibitem [{Note1()}]{Note1}%
  \BibitemOpen
  \bibinfo {note} {We verified that fluctuations in the phase boundaries due to
  finite-size effects become negligibly small for lattices larger than $36
  \times 36$ unit cells.}\BibitemShut {Stop}%
\bibitem [{Note2()}]{Note2}%
  \BibitemOpen
  \bibinfo {note} {We verified this statement by running unbiased MC
  simulations for multiple randomly chosen points within each phase on finite
  lattices of $12 \times 12$ unit cells.}\BibitemShut {Stop}%
\bibitem [{\citenamefont {Volovik}(2003)}]{Volovik2003}%
  \BibitemOpen
  \bibfield  {author} {\bibinfo {author} {\bibfnamefont {G.~E.}\ \bibnamefont
  {Volovik}},\ }\href@noop {} {\emph {\bibinfo {title} {The Universe in a
  Helium Droplet}}}\ (\bibinfo  {publisher} {Oxford University Press},\
  \bibinfo {address} {New York, USA},\ \bibinfo {year} {2003})\BibitemShut
  {NoStop}%
\bibitem [{\citenamefont {Hermanns}\ and\ \citenamefont
  {Trebst}(2014)}]{hermanns2014quantum}%
  \BibitemOpen
  \bibfield  {author} {\bibinfo {author} {\bibfnamefont {M.}~\bibnamefont
  {Hermanns}}\ and\ \bibinfo {author} {\bibfnamefont {S.}~\bibnamefont
  {Trebst}},\ }\href {\doibase 10.1103/PhysRevB.89.235102} {\bibfield
  {journal} {\bibinfo  {journal} {Phys. Rev. B}\ }\textbf {\bibinfo {volume}
  {89}},\ \bibinfo {pages} {235102} (\bibinfo {year} {2014})}\BibitemShut
  {NoStop}%
\bibitem [{\citenamefont {Hal\'asz}\ \emph {et~al.}(2016)\citenamefont
  {Hal\'asz}, \citenamefont {Perkins},\ and\ \citenamefont {van~den
  Brink}}]{Gabor2016}%
  \BibitemOpen
  \bibfield  {author} {\bibinfo {author} {\bibfnamefont {G.~B.}\ \bibnamefont
  {Hal\'asz}}, \bibinfo {author} {\bibfnamefont {N.~B.}\ \bibnamefont
  {Perkins}}, \ and\ \bibinfo {author} {\bibfnamefont {J.}~\bibnamefont
  {van~den Brink}},\ }\href {\doibase 10.1103/PhysRevLett.117.127203}
  {\bibfield  {journal} {\bibinfo  {journal} {Phys. Rev. Lett.}\ }\textbf
  {\bibinfo {volume} {117}},\ \bibinfo {pages} {127203} (\bibinfo {year}
  {2016})}\BibitemShut {NoStop}%
\bibitem [{\citenamefont {Hal\'asz}\ \emph {et~al.}(2017)\citenamefont
  {Hal\'asz}, \citenamefont {Perreault},\ and\ \citenamefont
  {Perkins}}]{Gabor2017}%
  \BibitemOpen
  \bibfield  {author} {\bibinfo {author} {\bibfnamefont {G.~B.}\ \bibnamefont
  {Hal\'asz}}, \bibinfo {author} {\bibfnamefont {B.}~\bibnamefont {Perreault}},
  \ and\ \bibinfo {author} {\bibfnamefont {N.~B.}\ \bibnamefont {Perkins}},\
  }\href {\doibase 10.1103/PhysRevLett.119.097202} {\bibfield  {journal}
  {\bibinfo  {journal} {Phys. Rev. Lett.}\ }\textbf {\bibinfo {volume} {119}},\
  \bibinfo {pages} {097202} (\bibinfo {year} {2017})}\BibitemShut {NoStop}%
\bibitem [{\citenamefont {Knolle}\ \emph {et~al.}(2014)\citenamefont {Knolle},
  \citenamefont {Kovrizhin}, \citenamefont {Chalker},\ and\ \citenamefont
  {Moessner}}]{knolle2014dynamics}%
  \BibitemOpen
  \bibfield  {author} {\bibinfo {author} {\bibfnamefont {J.}~\bibnamefont
  {Knolle}}, \bibinfo {author} {\bibfnamefont {D.~L.}\ \bibnamefont
  {Kovrizhin}}, \bibinfo {author} {\bibfnamefont {J.~T.}\ \bibnamefont
  {Chalker}}, \ and\ \bibinfo {author} {\bibfnamefont {R.}~\bibnamefont
  {Moessner}},\ }\href {\doibase 10.1103/PhysRevLett.112.207203} {\bibfield
  {journal} {\bibinfo  {journal} {Phys. Rev. Lett.}\ }\textbf {\bibinfo
  {volume} {112}},\ \bibinfo {pages} {207203} (\bibinfo {year}
  {2014})}\BibitemShut {NoStop}%
\bibitem [{\citenamefont {Knolle}\ \emph {et~al.}(2015)\citenamefont {Knolle},
  \citenamefont {Kovrizhin}, \citenamefont {Chalker},\ and\ \citenamefont
  {Moessner}}]{knolle2015dynamics}%
  \BibitemOpen
  \bibfield  {author} {\bibinfo {author} {\bibfnamefont {J.}~\bibnamefont
  {Knolle}}, \bibinfo {author} {\bibfnamefont {D.~L.}\ \bibnamefont
  {Kovrizhin}}, \bibinfo {author} {\bibfnamefont {J.~T.}\ \bibnamefont
  {Chalker}}, \ and\ \bibinfo {author} {\bibfnamefont {R.}~\bibnamefont
  {Moessner}},\ }\href {\doibase 10.1103/PhysRevB.92.115127} {\bibfield
  {journal} {\bibinfo  {journal} {Phys. Rev. B}\ }\textbf {\bibinfo {volume}
  {92}},\ \bibinfo {pages} {115127} (\bibinfo {year} {2015})}\BibitemShut
  {NoStop}%
\bibitem [{\citenamefont {Nasu}\ \emph {et~al.}(2014)\citenamefont {Nasu},
  \citenamefont {Udagawa},\ and\ \citenamefont {Motome}}]{Nasu2014}%
  \BibitemOpen
  \bibfield  {author} {\bibinfo {author} {\bibfnamefont {J.}~\bibnamefont
  {Nasu}}, \bibinfo {author} {\bibfnamefont {M.}~\bibnamefont {Udagawa}}, \
  and\ \bibinfo {author} {\bibfnamefont {Y.}~\bibnamefont {Motome}},\ }\href
  {\doibase 10.1103/PhysRevLett.113.197205} {\bibfield  {journal} {\bibinfo
  {journal} {Phys. Rev. Lett.}\ }\textbf {\bibinfo {volume} {113}},\ \bibinfo
  {pages} {197205} (\bibinfo {year} {2014})}\BibitemShut {NoStop}%
\bibitem [{\citenamefont {Nasu}\ \emph {et~al.}(2015)\citenamefont {Nasu},
  \citenamefont {Udagawa},\ and\ \citenamefont {Motome}}]{Nasu2015}%
  \BibitemOpen
  \bibfield  {author} {\bibinfo {author} {\bibfnamefont {J.}~\bibnamefont
  {Nasu}}, \bibinfo {author} {\bibfnamefont {M.}~\bibnamefont {Udagawa}}, \
  and\ \bibinfo {author} {\bibfnamefont {Y.}~\bibnamefont {Motome}},\ }\href
  {\doibase 10.1103/PhysRevB.92.115122} {\bibfield  {journal} {\bibinfo
  {journal} {Phys. Rev. B}\ }\textbf {\bibinfo {volume} {92}},\ \bibinfo
  {pages} {115122} (\bibinfo {year} {2015})}\BibitemShut {NoStop}%
\bibitem [{Note3()}]{Note3}%
  \BibitemOpen
  \bibinfo {note} {We average over 8 independent runs to estimate the
  errors.}\BibitemShut {Stop}%
\bibitem [{Note4()}]{Note4}%
  \BibitemOpen
  \bibinfo {note} {The height of the peak in $C(T)/L^2$ is proportional to the
  system volume, $L^2$, for first-order transitions and to $L^{\alpha /\nu }$
  for second-order transitions.}\BibitemShut {Stop}%
\bibitem [{\citenamefont {den Nijs}(1979)}]{Nijs1979}%
  \BibitemOpen
  \bibfield  {author} {\bibinfo {author} {\bibfnamefont {M.~P.~M.}\
  \bibnamefont {den Nijs}},\ }\href {\doibase 10.1088/0305-4470/12/10/030}
  {\bibfield  {journal} {\bibinfo  {journal} {Journal of Physics A:
  Mathematical and General}\ }\textbf {\bibinfo {volume} {12}},\ \bibinfo
  {pages} {1857} (\bibinfo {year} {1979})}\BibitemShut {NoStop}%
\end{thebibliography}
\end{document}